\documentclass[10pt, conference]{IEEEtran}
\usepackage{hyperref}

\ifCLASSINFOpdf
\else
\fi

\usepackage{lscape, subfigure, latexsym, amssymb, amsmath, algorithmic, multirow}
\usepackage[linesnumbered, vlined, ruled]{algorithm2e}
\usepackage[pdftex]{graphicx}
\usepackage{subfigure}
\usepackage{multirow}
\usepackage{mathtools, bbm, color}

\usepackage{amsmath,amssymb,amsthm,mathrsfs,amsfonts,dsfont}

\newtheorem{theorem}{Theorem}
\newtheorem{lemma}{Lemma}
\newtheorem{corollary}{Corollary}
\newtheorem{defn}{Definition}[section]

\begin{document}

\title{FDI: Quantifying Feature-based Data Inferability}

%%%mark 1-4
\author{
Shouling Ji$^{\dag, \ddag}$, Haiqin Weng$^\dag$, Yiming Wu$^\dag$, 
Qinming He$^\dag$, Raheem Beyah$^\ddag$, Ting Wang$^\sharp$ \\

$^\dag$ Zhejiang University \\
$^\ddag$ Georgia Institute of Technology\\
$^\natural$ Huazhong University of Science and Technology \\
$^\sharp$ Lehigh University \\

%sji@zju.edu.cn
%Shouling Ji%, Weiqing Li, Neil Z. Gong, Prateek Mittal, and Raheem Beyah \\
%%Georgia Institute of Technology \\
%%sji@gatech.edu
%%\IEEEauthorblockN{Shouling Ji\IEEEauthorrefmark{2},
%%Weiqing Li\IEEEauthorrefmark{2},
%%Jing (Selena) He\IEEEauthorrefmark{4}
%%Mudhakar Srivatsa\IEEEauthorrefmark{3}, and
%%Raheem Beyah\IEEEauthorrefmark{2}
%%%Montgomery Scott\IEEEauthorrefmark{3} and
%%%Eldon Tyrell\IEEEauthorrefmark{4}
%%}\\
%%\IEEEauthorblockA{\IEEEauthorrefmark{2}%School of Electrical and Computer Engineering\\
%%Georgia Institute of Technology,%\\%,
%%%Atlanta, Georgia 30332--0765, USA\\
%%%Email: \{sji, wli64\}@gatech.edu, rbeyah@ece.gatech.edu
%%}
}


\maketitle

\begin{abstract}
Motivated by many existing security and privacy applications,
e.g., network traffic attribution,
linkage attacks, private web search, and feature-based data de-anonymization,
in this paper, we study the Feature-based Data Inferability (FDI) quantification problem.
First, we conduct the FDI quantification under both naive and general
data models from both a feature distance perspective
and a feature distribution perspective.
Our quantification explicitly shows the conditions
to have a desired fraction of the target users to be
Top-$K$ inferable ($K$ is an integer parameter).
Then, based on our quantification, we evaluate the user inferability
in two cases: network traffic attribution in network forensics and
feature-based data de-anonymization.
Finally, based on the quantification and evaluation,
we discuss the implications of this research for existing
feature-based inference systems.
\end{abstract}

%\begin{IEEEkeywords}
%Classification, feature, machine learning, quantification, evaluation, network trace
%\end{IEEEkeywords}

% \fi
\IEEEpeerreviewmaketitle

\section{Introduction} \label{intro}

\begin{figure}[!tp]
 \centering
  %\subfigure[]{
    \includegraphics[width=3in]{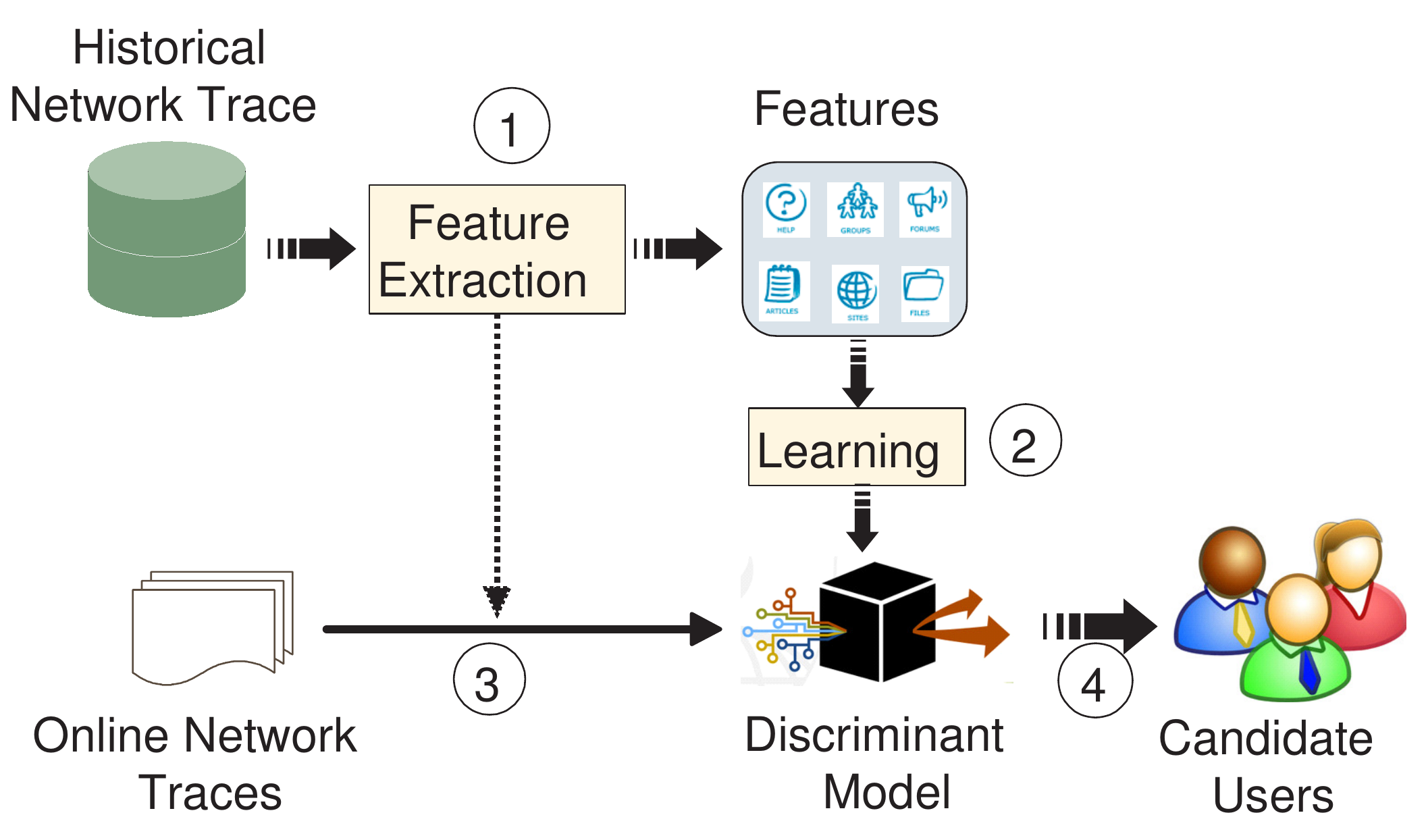}
  %}%\hspace{-8mm}
%  \subfigure[]{
%    \includegraphics[width=2.7in]{figures/traffic-eps-converted-to.pdf}
%  }%\hspace{-8mm}
 % \vspace{-3mm}
  \caption{Network traffic attribution system.} \label{f_example}
  %\vspace{-6mm}
%\vspace{-8mm}
\end{figure}

Many existing security and privacy applications/techniques
can be characterized
as a \emph{feature-based inference system},
e.g., network traffic attribution in network forensic applications,
private web search, feature-based data de-anonymization \cite{wanwansdm14}-\cite{narshmsp08}.
To conduct network traffic attribution, usually,
a network traffic attribution system is first learned based on
the features extracted from historical network traces.
Later, when new network traffic comes, features will be extracted
from the new traffic first, and then
the data will be automatically
attributed to the users who generated them
by the system based on the features (as shown in Fig.\ref{f_example})
\cite{wanwansdm14}. In fact, the network traffic attribution system
can be directly considered as a feature-based inference system,
where the system is first learned based on the historical/training data
(in detail, features of the historical/training) and
then used to infer the new data (in this scenario, users who generate
the new traffic) based on their features (as shown in Fig.\ref{f_model}).
Another example is the code stylometry-based de-anonymization attack
to programmers proposed in \cite{calharusenix15}.
In this kind of attack, the code stylometry features of training programs
are first extracted to train a de-anonymization model.
Then, this model can be used to de-anonymize the programmers of
the target programs based on their code stylometry features.
For this example, the code stylometry-based de-anonymization model
can also be considered as a feature-based inference system to infer (de-anonymize)
target data (programmers of targeting programs).

Now, some interesting questions are brought:
how to quantify the performance of those feature-based inference systems
for security and privacy applications? and
what is the performance of existing feature-based inference techniques
relative to the inherent theoretical performance bound?
Answering these questions are important
to accurately evaluate and understand the performance of existing
feature-based inference systems/techniques and further develop improved ones.
Unfortunately, although we already have many feature-based inference systems/techniques
for various security and privacy applications,
the answers to the brought questions remain unclear.
Therefore, to address these open problems,
in this paper, we study the Feature-based Data Inferability (FDI) quantification
for existing feature-based inference systems/techniques in various
security and privacy applications.
Particularly, we make the following contributions in this paper.
\begin{itemize}
\item
We first quantify the FDI under a naive data model,
where each user-feature relationship is characterized by a binary function
(a user either has a feature or does have a feature).
Under the naive model, we quantified the conditions to have
a target dataset to be $(\delta, K)$-inferable, i.e.,
to have $\delta \widetilde{m}$ target users to be Top-$K$ inferable,
where $\delta$ is a parameter in $[0, 1]$,
$\widetilde{m}$ is the number of overlapped users between the training data
of the inference model and the targeting data,
(thus, $\delta \widetilde{m}$ is the number of users that can be
correctly Top-$K$ inferred),
and $K$ is an integer specifying the desired inference accuracy.

\item
Subsequently, we extend our FDI quantification to a
general data model. Under the general data model,
we quantify the FDI from both the feature distance perspective
and the feature distribution perspective to have a target dataset
to be $(\delta, K)$-inferable.
Our quantification in the general scenarios provides the answers to
the raised open problems, and meanwhile, our quantification provides
the theoretical foundation for the first time for existing feature-based inference
systems in various security and privacy applications,
to the best of our knowledge.

\item
Based on our FDI quantification, we conduct a large-scale
evaluation leveraging on real world data.
Specifically, we evaluate the user inferability in two cases:
network traffic attribution in network forensics
and feature-based data de-anonymization.
We explicitly demonstrate the $(\delta, K)$-inferability
of users in these two cases and analyze the reasons.

\item
In terms of our quantification and evaluation,
we discuss the implications of this paper to practical
feature-based inference systems/techniques.
We also point out the future research directions.
\end{itemize}

The rest of this paper is organized as follows.
In Section \ref{problem}, we describe the motivation applications
and formalize the problem.
In Section \ref{quantification}, we quantify the FDI under both naive
and general data models.
In Section \ref{evaluation}, we evaluate the FDI in two scenarios.
We make further discussion in Section \ref{discussion}.
In Section \ref{related}, we summarize the related work and
we conclude the paper in Section \ref{conclusion}.

\section{Problem Formalization} \label{problem}

In this section, we formalize the studied problem.
To make the problem easily understandable and to further motivate our research,
we start from introducing motivation examples that
our study is applicable for analysis.

\subsection{Motivation Examples} \label{example}

%\subsubsection{Network Traffic Attribution.}

In this paper, we study data's feature-based inferability.
Our study is motivated by several existing security and privacy applications,
e.g., network traffic attribution in network security forensics
\cite{wanwansdm14}\cite{davkulicml07}\cite{neaperccs14},
linkage attacks and private web search \cite{gershoccs14}\cite{baltrosp12},
and data de-anonymization \cite{calharusenix15}\cite{afrcalsp14}\cite{narshmsp08}.

Network traffic attribution is one of the fundamental issues
in network security forensics,
under which users, who are responsible for the observed activities and behaviors
on network interfaces, are inferred \cite{wanwansdm14}\cite{neaperccs14}.
Taking the network traffic attribution system Kaleido proposed in \cite{wanwansdm14}
and shown in Fig.\ref{f_example} as an example, a typical
network traffic attribution system works as follows:
\textcircled{1}, based on the historical network traces,
a set of features (corresponding to each user) are extracted;
\textcircled{2}, a learning model is designed to learn
a discriminant model based on the features of historical network traces,
which is used for network traffic attribution and/or
new user (could be an intruder) identification;
\textcircled{3}, when new network traffic comes,
the features of the new network traffic are extracted;
and \textcircled{4}, taking the features of the new network traffic
as input, the discriminant model either attributes the traffic
to a set of candidate users or concludes that the traffic is generated
by a new user (a set of new users).

Web searching is one of the most fundamental computer applications,
by which users obtain desired knowledge and/or find interested websites.
Intuitively, users' web search traces carry users' interests and intents.
Therefore, potential adversaries (e.g., eavesdroppers) may design some linkage attacks
and exploit users' web search traces to infer users' profiles and other sensitive information
\cite{gershoccs14}\cite{baltrosp12}.
The key idea of a linkage attack is that ($i$) an adversary first
learns a linkage function based on the features of target users' historical
web search data and then ($ii$) determines whether the new generated
web search data/events belong to the target users.
To defend against the linkage attack in web search applications,
several obfuscation mechanisms have been proposed
for private web search \cite{gershoccs14}\cite{baltrosp12}.
The basic idea is to obfuscates users' web search data by adding
some noise, i.e., obfuscating the features of users' web search data
such that the linkage attack cannot effectively infer the generator of the data.

Our study in this paper is also motivated by existing
feature-based de-anonymization attacks and techniques,
e.g., programmers de-anonymization \cite{calharusenix15},
authorship distribution to underground forums
and multi-author detection \cite{afrcalsp14},
and movie rating data de-anonymization \cite{narshmsp08}.
In these de-anonymization attacks/techniques,
a feature-based de-anonymization model is first learned based on
a training dataset. Subsequently, the new coming data
(generated by an existing user or a new user) are de-anonymized
by the de-anonymization model based on the data's features.

\begin{figure}[!tp]
 \centering
    \includegraphics[width=3.6in]{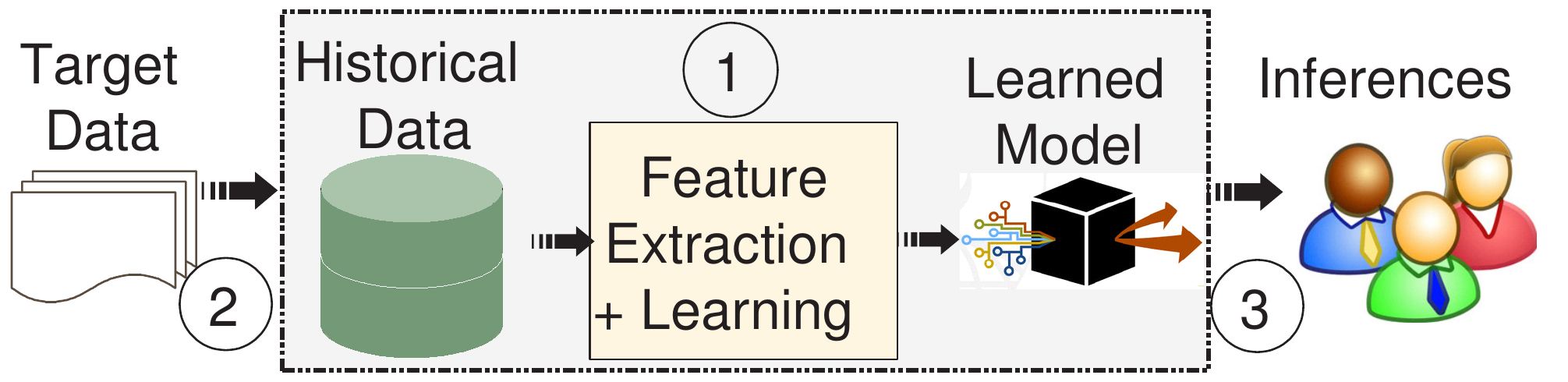}
  \caption{System architecture of the motivation applications.} \label{f_model}
\end{figure}

Mathematically, all the aforementioned security and privacy applications can be reduced
to a simple yet general system as shown in Fig.\ref{f_model}:
\textcircled{1}, a model is learned based on the features
of historical data; \textcircled{2}, the target data are input
to the model; and \textcircled{3},
inferences, e.g., candidate users who
generate the data and/or identified new users, are concluded
based on the results of the learned model.
Now, after observing the success of the aforementioned
security and privacy applications \cite{wanwansdm14}-\cite{narshmsp08}, e.g.,
Kaleido is able to identify the responsible users with
over $80\%$ accuracy,
two interesting questions are that why these techniques/attacks are success and
given the target data,
how to determine the performance of these techniques/attacks
relative to the \emph{intrinsic} inferability of the target data, e.g.,
how good the $80\%$ accuracy of Kaleido is and is that possible to achieve
some better accuracy than $80\%$? To answer the two questions, we study the
\emph{intrinsic inferability} of the target data given
the historical data (training data).
Therefore, our research in this paper can serve as the theoretical foundation
of the aforementioned security and privacy applications.
Furthermore, our quantification enables the development of a tool
to evaluate the relative performance of the aforementioned techniques/attacks
and guides the development of future research (as discussed in Section \ref{discussion}).

\subsection{Problem Formalization and Models}

Now, we formalize the studied problem.
During the formalization, the basic principle is to make the problem
sufficiently general and meanwhile mathematically tractable.

We denote the training data (e.g., the historical data
in the network traffic attribution scenario) as $U$.
Since we do not distinguish a user and the data generated by
that user, we assume $U$ consists of $n$ users (or the data generated by $n$ users),
and further assume $U = \{u_i | i = 1, 2, \cdots, n\}$,
where $u_i$ is a user (or the data generated by a user).
For $\forall u_i \in U$, it represents a user or the data generated by
a user depending on the context.
To model the feature extraction process (as shown in Fig.\ref{f_example} and
Fig.\ref{f_model}), we assume there is a \emph{feature extraction mechanism}
$\mathcal{F}  = \{f^1, f^2, \cdots, f^N\}$
\footnote{In practice, $\mathcal{F}$ could be any specific feature
extraction mechanism, e.g., the ones in \cite{wanwansdm14}-\cite{narshmsp08}.},
where $f^i$ denotes some particular feature function and
$N$ is the dimension of the feature space.
Applying $\mathcal{F}$ to $U$, we can get the features of $U$,
denoted by set $\mathcal{F} (U)$. In this paper, we focus on the scenario
that $\mathcal{F}(U)$ is a finite set,
i.e., $N$ is some finite value\footnote{With this assumption,
the studied problem is still sufficiently general to be applied to
many existing security and privacy applications. For instance,
in network security forensics
\cite{wanwansdm14}\cite{davkulicml07}\cite{neaperccs14},
linkage attacks and private web search \cite{gershoccs14}\cite{baltrosp12},
and data de-anonymization \cite{calharusenix15}\cite{afrcalsp14}\cite{narshmsp08},
the extracted features of the training data can be modeled by a finite set.}.
Specifically, for $\forall u_i \in U$, its features with respect to $\mathcal{F}$
are denoted by vector
$\overrightarrow{\mathcal{F}(u_i)} = <f_{u_i}^1, f_{u_i}^2, \cdots, f_{u_i}^N>$, where
$f_{u_i}^k$ $(1 \leq k \leq N)$ denotes the feature of $u_i$
with respect to the feature function $f^k \in \mathcal{F}$.

Similar to formalizing the training data and taking account of the security and privacy
applications (\cite{wanwansdm14}-\cite{narshmsp08}),
we denote the target data by $V = \{v_j | j = 1, 2, \cdots, m\}$,
where $v_j$ is a user (or the data generated by a target user) in the target data
and $m$ is the number of users in the target data.
As shown in Fig.\ref{f_example} and Fig.\ref{f_model} (\cite{wanwansdm14}-\cite{narshmsp08}),
before inferring the users in $V$, we apply the same $\mathcal{F}$ to extract the
features of $V$ denoted by $\mathcal{F}(V)$,
which is again assumed to be a finite set.
For $\forall v_j \in V$, its features with respect to $\mathcal{F}$
are denoted by vector $\overrightarrow{\mathcal{F}(v_j)} = <f^1_{v_j}, f^2_{v_j},
\cdots, f^N_{v_j}>$, where $f^k_{v_j}$ denotes the feature of $v_j$
with respect to the feature function $f^k \in \mathcal{F}$.
After having $\mathcal{F}(V)$, the task now is to infer
the users in $V$ using an inference model (e.g., the network traffic discriminant
model as shown in Fig.\ref{f_example}).

Based on the aforementioned definitions, the studied problem
in this paper can be formalized as follows:
\begin{defn}
Feature-based Data Inferability (FDI).
Given $U$, $V$, and $\mathcal{F}$, we quantify the inferability of
$V$ with respect to $U$ and $\mathcal{F}$.
\end{defn}
In this paper, we study the intrinsic FDI of the security and privacy
applications as shown Section \ref{example}.
Mathematically, the FDI study can serve as the theoretical foundation
of the applications in Section \ref{example},
e.g., the network traffic distribution system Kaleido
proposed in \cite{wanwansdm14}.
Practically, the FDI study can be employed to evaluate the relative
performance of the existing techniques in the applications of Section \ref{example},
and guide the development of new/improved techniques.

\section{FDI Quantification} \label{quantification}

In this section, we conduct the FDI quantification.
We start the quantification from a naive scenario.
Then, we generalize the FDI quantification to the more
practical cases.

To make our following discussion easily understandable,
we use the network traffic attribution application in network security forensics
as the studying context without of dedicated specification in the rest of this paper.
Straightforwardly, our discussion is applicable to the
scenarios of the linkage attack and private web search \cite{gershoccs14}\cite{baltrosp12}
and data de-anonymization \cite{calharusenix15}\cite{afrcalsp14}\cite{narshmsp08}.

\subsection{Preliminary}

Following the security and privacy applications in \cite{wanwansdm14}-\cite{narshmsp08},
an \emph{inferring model} can be learned from $\mathcal{F}(U)$ as shown in Fig.\ref{f_example}
and Fig.\ref{f_model},
e.g., the discriminant model in the network security forensics
application \cite{wanwansdm14}\cite{neaperccs14},
the linkage attack model in private web searching \cite{gershoccs14}\cite{baltrosp12},
and the de-anonymization model in \cite{calharusenix15}\cite{afrcalsp14}\cite{narshmsp08}.
We denote the inference (attack, de-anonymization) model by $\mathcal{M}$.
Then, $\mathcal{M}$ is employed to infer the new coming data,
i.e., the target data.

When employing $\mathcal{M}$ to infer users (data generated by users) in the target data,
$\mathcal{M}$ employs some inference function learned from $\mathcal{F}(U)$.
We here model the inference function of $\mathcal{M}$ by $\phi(\cdot, \cdot)$.
Then, $\forall v_j \in V$, when inferring $v_j$ using $\mathcal{M}$,
we denote the process by $\mathcal{M}(v_j: U)$ and
denote the inference result by
$\mathcal{M} (v_j: U) = \{u_i | u_i \in U, \phi(u_i, v_j)
\text{ returns true}\} \cup \{\triangle\}$, where
$\triangle$ denotes a new user (the data generated by a new user) such that
$\triangle \notin U$.
We further explain the inference result definition as follows:
when employing $\mathcal{M}$ to infer the target user (data generated by the target user)
$v_j$, it may be inferred to some candidate users in the training data $U$
if the inference function $\phi(\cdot, \cdot)$ is satisfied.
Otherwise, $\mathcal{M}$ is more confident to infer $v_j$ as a new user
that never appeared in $U$. For instance, in the network traffic
distribution application, when using Kaleido ($\mathcal{M}$ in our definition)
to monitor the on-line network traffic,
the inference result could be that the traffic is generated by some existing user
(used for training Kaleido) or the traffic is generated by some new user
that not appeared before (could be some intruder).
Now, we are ready to start our quantification.

\subsection{Warmup: Naive Quantification}

In this subsection, we conduct the FDI quantification for
a naive scenario, where we assume that
$\forall f^k \in \mathcal{F}$, $f^k$ is a \emph{binary}
feature function, i.e., $\forall u_i \in U$ or $\forall v_j \in V$,
$u_i$ or $v_j$ either has feature $f^k$ or not.
Then, we have $\forall w \in U \cup V$,
$\overrightarrow{\mathcal{F}(w)} = <f^k_{w} | f^k_{w} \in \{0, 1\}, k = 1, 2, \cdots, N>$,
i.e., the feature vector of $w$ is a $N$-dimensional 0-1 vector with respect to $\mathcal{F}$.
Furthermore, for $\overrightarrow{\mathcal{F}(w)}$, we define $\Gamma_w = \sum\limits_{k=1}^N f^k_w$.
Given two 0-1 vectors $\overrightarrow{\mathcal{F}(x)}$
and $\overrightarrow{\mathcal{F}(y)}$ where $x, y \in U \cup V$,
we define $\overrightarrow{\mathcal{F}(x \oplus y)} = <f_x^k \oplus f_y^k | k = 1, 2, \cdots, N>$,
where $\oplus$ is the logical binary XOR operation.

For $v \in V$ and $u \in U$, we denote $v \simeq u$ the scenario that
$v$ and $u$ correspond to the same user (or the data generated by the same user)
and $v \neq u$ otherwise, e.g., the network traffic generated
by the same user in different time windows or not.
To conduct the FDI quantification, the first step is to
understand and quantify the correlation of the features
of $v$ and $u$. Toward this objective,
for $v \in V$ and $u \in U$, we assume that
$\Pr(f_{v}^k = f_{u}^k | v \simeq u) = p$ for $1 \leq k \leq N$,
i.e., the probability that $v$ preserves the same property of $u$
with respect to a feature is $p$.
Now, for $u, w \in U$ and $v \in V$, suppose $v \simeq u$
while $v \neq w$. Then, we have the following lemma,
which quantifies the inferability of $v$ with respect to
$u$ and $w$\footnote{Note that, all the quantifications in this paper
are statistically meaningful, i.e., statistically, with probability of 1,
the FDI quantifications hold.}.

\begin{lemma} \label{l1}
If $p \neq 1/2$ and $\Gamma_{u \oplus w} \geq \frac{16 \ln N + 8}{(1 - 2p)^2}$,
then $\exists \mathcal{M}$ such that $\mathcal{M}(v : \{u, w\}) = \{u\}$,
i.e., $v$ is inferable with respect to $u$ and $w$.
\end{lemma}

\begin{figure}[!tp]
 \centering
    \includegraphics[width=2in]{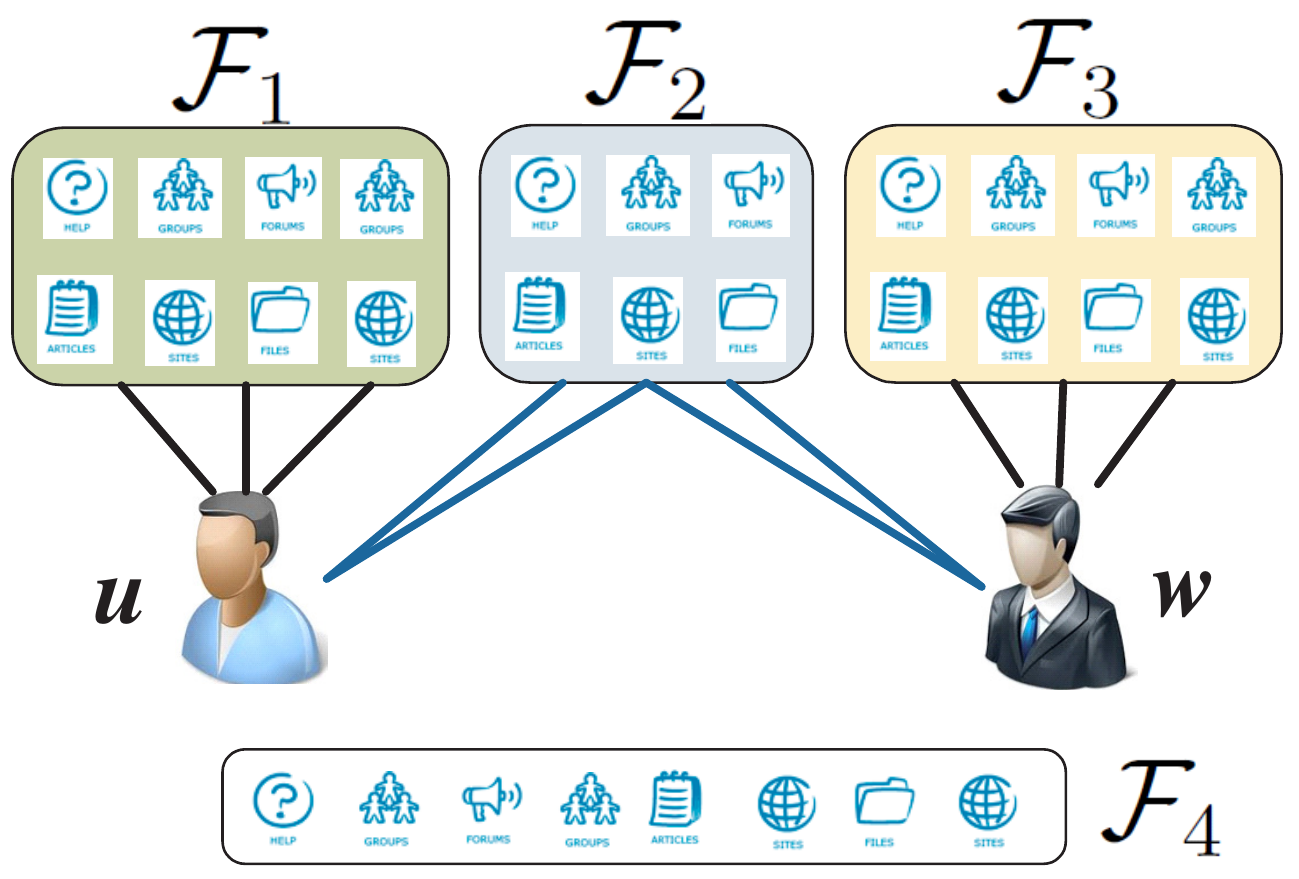}
  \caption{Feature space partition.} \label{f_attribute}
\end{figure}

{\em Proof}:
To prove this lemma, we first analyze the difference between
$\Gamma_{v \oplus w}$ and $\Gamma_{v \oplus u}$.
To facilitate our analysis, we partition the feature space $\mathcal{F}$
into four disjoint subsets with respect to $\mathcal{F}(u)$ and $\mathcal{F}(w)$,
denoted by $\mathcal{F}_1$, $\mathcal{F}_2$, $\mathcal{F}_3$, and $\mathcal{F}_4$ respectively
as shown in Fig.\ref{f_attribute},
where $\mathcal{F}_1 = \{f^k | f^k_u = 1, f^k_w = 0\}$ (the set of features
that $u$ has while $w$ dose not have),
$\mathcal{F}_2 = \{f^k | f^k_u = f^k_w = 1\}$ (the set of features that
both $u$ and $w$ have),
$\mathcal{F}_3 = \{f^k | f^k_u = 0, f^k_w = 1\}$ (the set of features that
$u$ does have while $w$ has),
and $\mathcal{F}_4 = \{f^k | f^k_u = f^k_w = 0\}$ (the set of features that
neither $u$ nor $w$ has).
Let $k_i = |\mathcal{F}_i|$ for $i = 1, 2, 3, 4$,
where $|\cdot|$ is the cardinality of a set.
Furthermore, for $x \in \{u, w, v\}$ and $1 \leq i \leq 4$,
let $\overrightarrow{\mathcal{F}_i(x)}$ be the
feature vector of $x$ with respect to the features in $\mathcal{F}_i$.
Evidently, $\overrightarrow{\mathcal{F}_i(x)}$ is a subvector of $\overrightarrow{\mathcal{F}(x)}$.
Furthermore, let $\Gamma^i_x = \sum\limits_{f^k \in \mathcal{F}_i} f^k_x$.
Then, it is easy to show that $\forall x, y \in \{u, w, v\}$,
$\Gamma_{x \oplus y} = \sum\limits_{i = 1}^4 \Gamma_{x \oplus y}^i$.

Let $\Lambda^v_{u, w} = \Gamma_{v \oplus w} - \Gamma_{v \oplus u}$.
Since $\Gamma_{v \oplus w} = \sum\limits_{i = 1}^4 \Gamma_{v \oplus w}^i$
and $\Gamma_{v \oplus u} = \sum\limits_{i = 1}^4 \Gamma_{v \oplus u}^i$,
we have $\Lambda^v_{u, w} = \sum\limits_{i = 1}^4 (\Gamma_{v \oplus w}^i - \Gamma_{v \oplus u}^i)$.
Now, we consider each $\mathcal{F}_i$ separately:
(1) since both $u$ and $w$ have the features
in $\mathcal{F}_2$, we have $\Gamma_{v \oplus w}^2 - \Gamma_{v \oplus u}^2 = 0$;
(2) similar to $\mathcal{F}_2$, since neither $u$ nor $w$
has any feature in $\mathcal{F}_4$, we have $\Gamma_{v \oplus w}^4 - \Gamma_{v \oplus u}^4 = 0$;
(3) for $\mathcal{F}_1$, the set of features hold by $u$ while not $w$,
statistically, we have $\Gamma_{v \oplus u}^1 \sim B(k_1, 1 - p)$
and $\Gamma_{v \oplus w}^1 \sim B(k_1, p)$, where $B(x, y)$ is a
\emph{binomial} variable with parameters $x$ and $y$;
and (4) for $\Gamma_{v \oplus w}^3$, the set of features hold by $w$ while not $u$,
statistically, we have $\Gamma_{v \oplus u}^3 \sim B(k_3, 1 - p)$
and $\Gamma_{v \oplus w}^3 \sim B(k_3, p)$.
Then, we have
\begin{align*}
& \Lambda^v_{u, w} \\
& = \sum\limits_{i = 1, 3} (\Gamma_{v \oplus w}^i - \Gamma_{v \oplus u}^i) \\
& \stackrel{statistically}{=} B(k_1, p) + B(k_3, p) - B(k_1, 1- p) - B(k_3, 1-p) \\
& = B(k_1 + k_3, p) - B(k_1 + k_3, 1 - p) \\
& = B(\Gamma_{u \oplus w}, p) - B(\Gamma_{u \oplus w}, 1 - p).
\end{align*}

Now, we consider two cases.
First, if $p > \frac{1}{2}$, we have $p\Gamma_{u \oplus w} > (1 - p) \Gamma_{u \oplus w}$.
Then, applying the Pedarsani-Grossglauser lemma \cite{pedgrokdd11},
we have
\begin{align*}
&\Pr(\Lambda^v_{u, w} \leq 0) \\
& \stackrel{statistically}{=} \Pr(B(\Gamma_{u \oplus w}, p) - B(\Gamma_{u \oplus w}, 1 - p) \leq 0) \\
& \leq 2 \exp(- \frac{(p\Gamma_{u \oplus w} - (1 - p) \Gamma_{u \oplus w})^2}
{8(p\Gamma_{u \oplus w} + (1 - p) \Gamma_{u \oplus w})}) \\
& = 2 \exp(- \frac{(2p - 1)^2 \Gamma_{u \oplus w}}{8}).
\end{align*}
Since $\Gamma_{u \oplus w} \geq \frac{16 \ln N + 8}{(1 - 2p)^2}$, we have
\begin{align*}
\Pr(\Lambda^v_{u, w} \leq 0)
& \stackrel{statistically}{\leq} 2\exp(- 2\ln N - 1) \\
& \leq \frac{1}{N^2}.
\end{align*}
Then, according to the Borel-Cantelli Lemma and statistically,
we have $\Pr(\Lambda^v_{u, w} \leq 0)\stackrel{N \rightarrow \infty}{=} 0$,
which implies that statistically,
$\Pr(\Gamma_{v \oplus w} > \Gamma_{v \oplus u}) \stackrel{N \rightarrow \infty}{=} 1$.

Second, we consider the case that $p < \frac{1}{2}$.
In this case, we have $p\Gamma_{u \oplus w} < (1 - p) \Gamma_{u \oplus w}$.
Then, applying the Pedarsani-Grossglauser lemma \cite{pedgrokdd11},
we have
\begin{align*}
&\Pr(\Lambda^v_{u, w} \geq 0) \\
& \stackrel{statistically}{=} \Pr(B(\Gamma_{u \oplus w}, p) - B(\Gamma_{u \oplus w}, 1 - p) \geq 0) \\
& \leq 2 \exp(- \frac{((1 - p) \Gamma_{u \oplus w} - p\Gamma_{u \oplus w})^2}
{8((1 - p) \Gamma_{u \oplus w} + p\Gamma_{u \oplus w})}) \\
& = 2 \exp(- \frac{(1 - 2p)^2 \Gamma_{u \oplus w}} {8}).
\end{align*}
Considering that $\Gamma_{u \oplus w} \geq \frac{16 \ln N + 8}{(1 - 2p)^2}$, we have
\begin{align*}
\Pr(\Lambda^v_{u, w} \geq 0)
& \stackrel{statistically}{\leq} 2\exp(- 2\ln N - 1) \\
& \leq \frac{1}{N^2}.
\end{align*}
According to the Borel-Cantelli Lemma and statistically,
we have $\Pr(\Lambda^v_{u, w} \geq 0)\stackrel{N \rightarrow \infty}{=} 0$,
i.e.,
$\Pr(\Gamma_{v \oplus w} < \Gamma_{v \oplus u}) \stackrel{N \rightarrow \infty}{=} 1$.

\begin{algorithm} \label{a_naive}
\SetKwInOut{Input}{input}
\SetKwInOut{Output}{output}
%\Input{$G^a, G^u, \mathcal{M}_s$}
%\Output{DA of $G^a$}
%\BlankLine
\If{$p > \frac{1}{2}$}
{
   $\mathcal{M}(v : \{u, w\} = \arg\min\{x | \Gamma_{v \oplus x}, x \in \{u, w\}\}$\;
}
\ElseIf{$p < \frac{1}{2}$}
{
    $\mathcal{M}(v : \{u, w\} = \arg\max\{x | \Gamma_{v \oplus x}, x \in \{u, w\}\}$\;
}
\caption{A naive implementation of $\mathcal{M}$.}
\end{algorithm}

Now, we need to show that $\exists \mathcal{M}$ such that $\mathcal{M}(v : \{u, w\}) = \{u\}$.
Based on our proof, it is trivial to show that
(1) when $p > \frac{1}{2}$, $\Pr(\mathcal{M}(v : \{u, w\}) = \{u\}) \stackrel{statistically}{=} 1$
if $\mathcal{M}$ is an \emph{increasing} function with respect to $\Gamma_{v \oplus x}$,
where $x \in \{u, w\}$; and similarly,
when $p < \frac{1}{2}$, $\Pr(\mathcal{M}(v : \{u, w\}) = \{u\}) \stackrel{statistically}{=} 1$
if $\mathcal{M}$ is a \emph{decreasing} function with respect to $\Gamma_{v \oplus x}$.
Therefore, for our purpose it is easy to design $\mathcal{M}$ using existing techniques
\cite{wanwansdm14}-\cite{narshmsp08}. To name a naive one, we can set $\mathcal{M}$
as shown in Algorithm \ref{a_naive}.
\hfill $\Box$

In Lemma \ref{l1}, we quantified the condition to successfully
infer user $v$ from $V$ with respect to $\{u, w\} \subseteq U$.
We further discuss Lemma \ref{l1} as follows.
First, one condition is that $p \neq \frac{1}{2}$.
This is consistent with our institution. If $p = \frac{1}{2}$,
the features of each user is uniformly and equiprobably
distributed in $\mathcal{F}$. Then, theoretically, all the users are equivalent
with respect to $\mathcal{F}$ and thus it is difficult (if not impossible)
to successfully infer $v \in V$ based on the features in $\mathcal{F}$
by any model.
Second, when $p \neq \frac{1}{2}$, we explicitly specify the condition
that $v \in V$ is statistically guaranteed to be successfully inferrable
with respect to $\{u, w\}$. In our proof, we also give how to design $\mathcal{M}$.
Note that, the specified condition is sufficient while not necessary
to have $v$ inferable with respect to $\{u, w\}$.
Even if the condition is not satisfied, it is also possible
to successfully infer $v$ with respect to $\{u, w\}$.
Particularly, we show this fact in the following corollary.

\begin{corollary} \label{c1}
For $v \in V$ and $u, w\in U$, suppose $v \simeq u$ and $v \neq w$.
If $p \neq 1/2$, then $\exists \mathcal{M}$ such that
$\Pr(\mathcal{M}(v : \{u, w\}) = \{u\}) \stackrel{statistically}{\geq}
\max\{0, 1 - 2 \exp(- \frac{(1 - 2p)^2 \Gamma_{u \oplus w}}{8})\}$.
\end{corollary}

{\em Proof}:
This corollary can be proven using the similar technique as
in Lemma \ref{l1}.
\hfill $\Box$

In Lemma \ref{l1}, we quantify the FDI of $v$ with respect to
$\{u, w\}$. Now, we quantify the FDI of $v$ with respect to $U$.
In practice, we usually infer $v$ to a set of candidate users in $U$.
For instance, in the network traffic distribution system Kaleido \cite{wanwansdm14},
the user responsible for the new coming traffic might be inferred
to a set of $K$ $(K \in [1, n])$ users.
Therefore, given $v \in V$, we define the Top-$K$ candidate set of $v$ as follows.

\begin{defn}
Top-$K$ candidate set and Top-$K$ inferable. For $v \in V$,
suppose that $\exists u \in U$ such that $u \simeq v$.
Then, the Top-$K$ candidate set of $v$ $(K \in [1, n])$, denoted by $\mathcal{K}_v$,
is defined as $\mathcal{K}_v \subseteq U$ such that $|\mathcal{K}_v| = K$
and $u \in \mathcal{K}_v$.
$v$ is Top-$K$ inferable with respect to $U$ if $\exists \mathcal{M}$
such that $\mathcal{M}(v : U) = \mathcal{K}_v$, i.e.,
$\mathcal{M}$ returns a subset of $U$ with size $K$ and $u$
is in that subset.
\end{defn}

Now, we quantify the Top-$K$ FDI of a user $v \in V$.
Let $\overline{\mathcal{K}_v}$ be a subset of $U$ such that $|\overline{\mathcal{K}_v}| = n - K$
and $v \notin \overline{\mathcal{K}_v}$.
We show the result in the following lemma.

\begin{lemma} \label{l2}
For $v \in V$, suppose that $v \simeq u \in U$.
Then, $v$ is Top-$K$ inferable if $p \neq 1/2$ and
$\exists \overline{\mathcal{K}_v} \subseteq U$ such that
$\min\{\Gamma_{u \oplus w} | w \in \overline{\mathcal{K}_v}\}
\geq \frac{16 \ln N + 8 \ln (2\theta n)}{(1 - 2p)^2}$,
where $\theta = \frac{n - K}{n}$.
\end{lemma}

{\em Proof}:
we prove this lemma by considering two cases.
First, we consider the case that $p > \frac{1}{2}$.
We define an event $E_1$ as $\exists w \in \overline{\mathcal{K}_v}$
such that $\Gamma_{v \oplus u} \geq \Gamma_{v \oplus w}$.
Then, we have $\Pr(E_1) = \Pr(\bigcup\limits_{w \in \overline{\mathcal{K}_v}}
\Gamma_{v \oplus u} \geq \Gamma_{v \oplus w})
\leq \sum\limits_{w \in \overline{\mathcal{K}_v}}
\Pr(\Gamma_{v \oplus u} \geq \Gamma_{v \oplus w})$ according to Boole's inequality.
From Lemma \ref{l1}, when $p>\frac{1}{2}$,
$\Pr(\Gamma_{v \oplus u} \geq \Gamma_{v \oplus w}) \stackrel{statistically}{\leq}
2 \exp(- \frac{(2p - 1)^2 \Gamma_{u \oplus w}}{8})$.
Then, we have
\begin{align*}
\Pr(E_1) & \stackrel{statistically}{\leq} \sum\limits_{w \in \overline{\mathcal{K}_v}}
2 \exp(- \frac{(2p - 1)^2 \Gamma_{u \oplus w}}{8}) \\
& \leq \sum\limits_{w \in \overline{\mathcal{K}_v}} 2 \exp(- 2 \ln N - \ln 2\theta n) \\
& = 2\theta n \exp(- 2 \ln N - \ln 2\theta n)
 = 1/ N^2.
\end{align*}
According to the Borel-Cantelli Lemma,
we have $\Pr(E_1) \stackrel{N \rightarrow \infty} = 0$,
i.e., $\Pr(\forall w \in \overline{\mathcal{K}_v},
\Gamma_{v \oplus u} < \Gamma_{v \oplus w}) \stackrel{N \rightarrow \infty}{=} 1$.

Second, we consider the case that $p < \frac{1}{2}$.
In this case, we define $E_2$ as an event that
$\exists w \in \overline{\mathcal{K}_v}$ such that $\Gamma_{v \oplus u}
\leq \Gamma_{v \oplus w}$.
Then, similar to the case that $p > \frac{1}{2}$, we have
\begin{align*}
\Pr(E_2)
& = \Pr(\bigcup\limits_{w \in \overline{\mathcal{K}_v}}
\Gamma_{v \oplus u} \leq \Gamma_{v \oplus w}) \\
& \leq \sum\limits_{w \in \overline{\mathcal{K}_v}}
\Pr(\Gamma_{v \oplus u} \leq \Gamma_{v \oplus w}) \\
& \leq \sum\limits_{w \in \overline{\mathcal{K}_v}}
2 \exp(- \frac{(1 - 2p)^2 \Gamma_{u \oplus w}} {8}) \\
& \leq 1/N^2.
\end{align*}
Again, according to the Borel-Cantelli Lemma,
we have $\Pr(E_2) \stackrel{N \rightarrow \infty} = 0$,
i.e., $\Pr(\forall w \in \overline{\mathcal{K}_v},
\Gamma_{v \oplus u} > \Gamma_{v \oplus w}) \stackrel{N \rightarrow \infty}{=} 1$.

\begin{algorithm} \label{a_topnaive}
\SetKwInOut{Input}{input}
\SetKwInOut{Output}{output}
%\Input{$G^a, G^u, \mathcal{M}_s$}
%\Output{DA of $G^a$}
%\BlankLine
$\mathcal{K}_v \leftarrow \{u_1, u_2, \cdots, u_K\}$\;
$U' \leftarrow U \setminus \mathcal{K}_v$\;
\If{$p > \frac{1}{2}$}
{
   \For{$x \in U'$}
   {
    $u' = \arg\max\limits_{y}\{\Gamma_{v\oplus y} | y \in \mathcal{K}_v\}$\;
    \If{$\Gamma_{v \oplus u'} > \Gamma_{v \oplus x}$}
    {
        $\mathcal{K}_v \leftarrow \mathcal{K}_v \setminus \{u'\}$\;
        $\mathcal{K}_v \leftarrow \mathcal{K}_v \cup \{x\}$\;
    }
   }
}
\ElseIf{$p < \frac{1}{2}$}
{
    \For{$x \in U'$}
   {
    $u' = \arg\min\limits_{y}\{\Gamma_{v\oplus y} | y \in \mathcal{K}_v\}$\;
    \If{$\Gamma_{v \oplus u'} < \Gamma_{v \oplus x}$}
    {
        $\mathcal{K}_v \leftarrow \mathcal{K}_v \setminus \{u'\}$\;
        $\mathcal{K}_v \leftarrow \mathcal{K}_v \cup \{x\}$\;
    }
   }
}
return $\mathcal{K}_v$\;
\caption{An implementation of $\mathcal{M}$ to have $v$ Top-$K$ inferable.}
\end{algorithm}

Now, we discuss how to design $\mathcal{M}$ and how to find
$\mathcal{K}_v$. Based on our proof,
if $p \neq 1/2$ and $\exists \overline{\mathcal{K}_v} \subseteq U$ such that
$\min\{\Gamma_{u \oplus w} | w \in \overline{\mathcal{K}_v}\}
\geq \frac{16 \ln N + 8 \ln (2\theta n)}{(1 - 2p)^2}$,
then (1) when $p > 1/2$,  $\Pr(\forall w \in \overline{\mathcal{K}_v},
\Gamma_{v \oplus u} < \Gamma_{v \oplus w}) \stackrel{N \rightarrow \infty}{=} 1$,
which implies that among $U$, there are at least $n - K$ users
having their $\Gamma_{v \oplus \cdot}$ values greater than
$\Gamma_{v \oplus u}$;
and (2) when $p < 1/2$, $\Pr(\forall w \in \overline{\mathcal{K}_v},
\Gamma_{v \oplus u} > \Gamma_{v \oplus w}) \stackrel{N \rightarrow \infty}{=} 1$,
there are at least $n - K$ users having their $\Gamma_{v \oplus \cdot}$ values
smaller than $\Gamma_{v \oplus u}$.
According to this observation, we give a preliminary implementation of $\mathcal{M}$
as shown in Algorithm \ref{a_topnaive}.
Basically, if $p > 1/2$, Algorithm \ref{a_topnaive} returns a set $\mathcal{K}_v$
consisting of $K$ users from $U$ that have the top-$K$ minimum
$\Gamma_{v \oplus \cdot}$ values;
and if $p < 1/2$, Algorithm \ref{a_topnaive} returns a set $\mathcal{K}_v$
consisting of $K$ users from $U$ that have the top-$K$ maximum
$\Gamma_{v \oplus \cdot}$ values.
By a contradiction-based technique, we can show that
the $\mathcal{M}$ shown in Algorithm \ref{a_topnaive}
returns a Top-$K$ candidate set of $v$,
i.e., $v$ is Top-$K$ inferable.
\hfill $\Box$

In Lemma \ref{l2}, the conditions for a user to be Top-$K$
inferable are quantified. If the specified conditions are satisfied,
we also provide an implementation of $\mathcal{M}$ in the proof (Algorithm \ref{a_topnaive}).
In fact, there are also many other techniques to implement $\mathcal{M}$,
e.g., the techniques proposed in \cite{wanwansdm14}-\cite{narshmsp08}.
Further, similar to Lemma \ref{l1}, the conditions in Lemma \ref{l2} are sufficient
while not necessary for $v$ to be Top-$K$ inferable.
When the conditions are satisfied, it is statistically guaranteed that $v$
is Top-$K$ inferable. Otherwise, $v$ is still Top-$K$ inferable with
some probability. Particularly, we show that probability in the following corollary.

\begin{corollary} \label{c2}
For $v \in V$, suppose that $v \simeq u \in U$.
Then, if $p \neq 1/2$, $\Pr(\mathcal{M}(v : U) = \mathcal{K}_v) \geq \max\{0,
1 - 2\theta n \exp(-\frac{(1 - 2p)^2 \Gamma_{\min} (\overline{\mathcal{K}_v})}{8})\}$,
where $\theta = \frac{n - K}{n}$ and $\Gamma_{\min} (\overline{\mathcal{K}_v})
= \min\{\Gamma_{u \oplus w} | w \in \overline{\mathcal{K}_v}\}$.
\end{corollary}

Now, we consider an even more general scenario where we try to infer multiple users
in $V$. A practical application corresponding to this scenario is to distribute the
monitored network traffic generated by multiple users in network forensics
\cite{wanwansdm14}\cite{neaperccs14}.
Let $\widetilde{V} = \{x | x \in V, \text{ and } \exists y \in U, s.t. \ x \simeq y\}$,
i.e., $\widetilde{V}$ is a set of users that appeared in both $V$ and $U$.
Furthermore, let $\delta$ be a constant and $\delta \in [0, 1]$.
Then, we define the $(\delta, K)$-inferability of $V$ (i.e.,
$V$ is $(\delta, k)$-inferable) as follows.

\begin{defn}
$(\delta, K)$-Inferable. $V$ is $(\delta, K)$-inferable if
there are at least $\delta \cdot \widetilde{m}$ users in $\widetilde{V}$
are Top-$K$ inferable\footnote{Without loss of generality,
we assume $\delta \widetilde{m}$ is an integer in $[0, m]$.
In the case that $\delta \widetilde{m}$ is not an integer,
we can define $\delta \widetilde{m}$ as $\delta \widetilde{m}
\leftarrow \lfloor\delta \widetilde{m}\rfloor$}.
\end{defn}

Then, we quantify the $(\delta, K)$-inferability of $V$ in the following
theorem.

\begin{theorem} \label{t1}
Let $V_\delta$ be any subset of $\widetilde{V}$
and $|V_\delta| = \delta \widetilde{m}$.
$V$ is $(\delta, K)$-inferable if
%($i$)
$p \neq 1/2$ and
%($ii$)
%$\exists V_\delta \subseteq \widetilde{V}$such that
%$|V_\delta| = \delta \widetilde{m}$ and
$\forall v \in V_\delta$,
$\exists \overline{\mathcal{K}_v} \subseteq U$
such that $|\overline{\mathcal{K}_v}| = n - K$,
and $\min\{\Gamma_{u \oplus w} | u \in U, u \simeq v, \text{ and } w \in
\overline{\mathcal{K}_v}\} \geq \frac{16 \ln N + 8\ln (2\delta \theta \widetilde{m}n)}{(1-2p)^2}$.
\end{theorem}

{\em Proof}:
We first prove this theorem for the case that $p > 1/2$.
For $v \in V_\delta$, suppose $v \simeq u \in U$.
Evidently, $|\mathcal{K}_v| = K$.
Now, to prove this theorem, it is sufficient to show that
$\forall v \in V_\delta$, $v$ is Top-$K$ inferable.
Let $E$ be the event that
$\exists v \in V_\delta$ such that $v$ is not Top-$K$ inferable.
Then, we have
\begin{align*}
\Pr(E)
& = \Pr(\bigcup\limits_{v \in V_\delta} v \text{ is not Top-}K \text{ inferable}) \\
& \leq \sum\limits_{v \in V_\delta} \Pr(v \text{ is not Top-}K \text{ inferable}) \\
& \leq \sum\limits_{v \in V_\delta} (1 - \Pr(\forall w \in \overline{\mathcal{K}_v},
\Gamma_{v \oplus u} < \Gamma_{v \oplus w})) \\
& = \sum\limits_{v \in V_\delta} \Pr(\exists w \in \overline{\mathcal{K}_v},
\Gamma_{v \oplus u} \geq \Gamma_{v \oplus w}) \\
\end{align*}
Then, according to Lemma \ref{l1} and Lemma \ref{l2},
we have
\begin{align*}
\Pr(E)
& \stackrel{statistically}{\leq} \sum\limits_{v \in V_\delta}
 \sum\limits_{w \in \overline{\mathcal{K}_v}}
2 \exp(- \frac{(2p - 1)^2 \Gamma_{u \oplus w}}{8}) \\
& \leq \sum\limits_{v \in V_\delta}
 \sum\limits_{w \in \overline{\mathcal{K}_v}}
 2 \exp(-2 \ln N - \ln (2 \delta \theta \widetilde{m} n)) \\
& = \sum\limits_{v \in V_\delta} \exp(\ln 2\theta n -
2 \ln N - \ln (2 \delta \theta \widetilde{m} n)) \\
& = 1/N^2.
\end{align*}
Following the Borel-Cantelli Lemma,
we have $\Pr(E) \stackrel{N \rightarrow \infty} = 0$,
i.e., $\Pr(\forall v \in V_\delta, v$ is Top-$K$ inferable$)
\stackrel{N \rightarrow \infty} = 1$ which implies that
$V$ is $(\delta, K)$-inferable.

For the case that $p < 1/2$, we have
\begin{align*}
\Pr(E)
& = \Pr(\bigcup\limits_{v \in V_\delta} v \text{ is not Top-}K \text{ inferable}) \\
& \leq \sum\limits_{v \in V_\delta} \Pr(v \text{ is not Top-}K \text{ inferable}) \\
& \leq \sum\limits_{v \in V_\delta} (1 - \Pr(\forall w \in \overline{\mathcal{K}_v},
\Gamma_{v \oplus u} > \Gamma_{v \oplus w})) \\
& = \sum\limits_{v \in V_\delta} \Pr(\exists w \in \overline{\mathcal{K}_v},
\Gamma_{v \oplus u} \leq \Gamma_{v \oplus w}).
\end{align*}
Then, according to Lemma \ref{l1} and Lemma \ref{l2},
we have
\begin{align*}
\Pr(E)
& \stackrel{statistically}{\leq} \sum\limits_{v \in V_\delta}
 \sum\limits_{w \in \overline{\mathcal{K}_v}}
2 \exp(- \frac{(1 - 2p)^2 \Gamma_{u \oplus w}}{8}) \\
%& \leq \sum\limits_{v \in V_\delta}
% \sum\limits_{w \in \overline{\mathcal{K}_v}}
% 2 \exp(-2 \ln N - \ln (2 \delta \theta \widetilde{m} n)) \\
%& = \sum\limits_{v \in V_\delta} \exp(\ln 2\theta n -
%2 \ln N - \ln (2 \delta \theta \widetilde{m} n)) \\
& \leq 1/N^2.
\end{align*}
Again, following the Borel-Cantelli Lemma,
we have $\Pr(E) \stackrel{N \rightarrow \infty} = 0$,
which implies that
$V$ is $(\delta, K)$-inferable.
\hfill $\Box$

In Theorem \ref{t1}, we quantify the $(\delta, K)$-inferability
of $V$. When comparing Theorem \ref{t1}
and Lemma \ref{l2}, we can see that
the conditions specified in Theorem \ref{t1} is
stronger than that in Lemma \ref{l2} with respect two aspects.
First, in Theorem \ref{t1}, it is required that for $\forall v \in V_\delta$,
there exists one desired $\overline{\mathcal{K}_v}$.
This is for the purpose of making $v$ Top-$K$ inferable.
Second, the required $\min{\Gamma_{u \oplus w}}$ is stronger
in Theorem \ref{t1} than that in Lemma \ref{l2}.
This can be explained from the statistical perspective.
In Lemma \ref{l2}, the objective is make one user statistically Top-$K$ inferable
while in Theorem \ref{t1}, the objective is make all the users
in $V_\delta$ statistically Top-$K$ inferable (simultaneously).

If the specified conditions in Theorem \ref{t1} are satisfied,
an interesting question is how to design a $\mathcal{M}$ to make
$V$ $(\delta, K)$-inferable. An preliminary implementation of $\mathcal{M}$
can be built using the procedure in Algorithm \ref{a_topnaive}:
for each user $v$ in $V$, we use Algorithm \ref{a_topnaive} to find
a $\mathcal{K}_v$ for it. Then, by the similar argument as in
Lemma \ref{l2}, we can conclude that $V$ is $(\delta, K)$-inferable
under $\mathcal{M}$.

In this subsection, we conduct the FDI quantification under the assumption
that each feature function is binary.
Apparently, this assumption may not hold in many real applications.
Nevertheless, the quantification in this subsection can shed light on
sophisticated FDI analysis. In the following subsections,
we consider general FDI quantification by removing this assumption.

\subsection{General Quantification: From the Distance Perspective} \label{dispers}

In the previous FDI quantification, we assume that $\forall f^i \in \mathcal{F}$,
$f^i$ is a binary function, i.e., $f^i \in \{0, 1\}$.
Although this assumption holds in many real applications
(e.g., linkage attacks
and data de-anonymization attacks),
$f^i$ may not be a binary function in many other applications.
Therefore, in the following FDI quantification, we assume
that $f^i$ can be any function with a real-value output.
Furthermore, given $\mathcal{F} = \{f^1, f^2, \cdots, f^N\}$,
an inference model $\mathcal{M}$ may assign different weights
to each feature (usually, the weights are learned from
the features of the training data, i.e., $\mathcal{F}(U)$).
To characterize this situation, we model that each feature $f^i$ in $\mathcal{F}$
corresponds to a weight value in $\mathcal{M}$,
which can be obtained by a weight function $w^i$.
In addition, to make our FDI quantification sufficiently general
and meanwhile mathematically tractable,
we model the correlation between the feature function $f^i$
and the weight function $w^i$ by another function
$g(f^i, w^i)$, i.e., $g(\cdot, \cdot)$ is a function defined
on $f^i$ and $w^i$ \footnote{Here, to make our model sufficiently general,
we do not specify the dedicated definition of $g(\cdot, \cdot)$.
In a specifical application, $g(\cdot, \cdot)$ can be specified accordingly.
For instance, we may have $g(f^i, w^i) = w^i \cdot f^i$ as in a
linear regression model.}.
Now, for a user $x \in V$ (or $x \in U$), we have
its feature vector as $\overrightarrow{\mathcal{F}(x)} = <g(f^i_x, w^i_x)
| 1 \leq i \leq N>$, where $g(f^i_x, w^i_x)$ is the function
defined on the feature function $f^i_x$ and the weight function $w^i_x$
of $x$.

Given $\mathcal{M}$ learned from $U$, we quantify the FDI of $V$
using $\mathcal{M}$. For instance, $V$ could be the new monitored network traffic
or the new collected web search data.
For $v \in V$, to infer $v$ to some user in $U$ (or the data in $U$ generated by the
same user) or to determine whether $v$ is a new user (or the data generated by a
new user), two fundamental approaches are usually employed in $\mathcal{M}$:
\emph{distance-based approach} and \emph{distribution-based approach}
\cite{wanwansdm14}-\cite{narshmsp08}.
In the distance-based approach, $\mathcal{M}$ computes the feature distance
between $v$ and each $u$ in $U$,
i.e., the distance between $\overrightarrow{\mathcal{F}(v)}$
and $\overrightarrow{\mathcal{F}(u)}$ for $u \in U$.
Then, $\mathcal{M}$ infers $v$ to a subset of candidates in $U$
(either has the minimum or the maximum distance value).
In the distribution-based approach, $\mathcal{M}$ computes the feature
distribution similarity between $v$ and each $u$ in $U$,
i.e., the distribution similarity between $\overrightarrow{\mathcal{F}(v)}$
and $\overrightarrow{\mathcal{F}(u)}$ for $u \in U$.
Then, $\mathcal{M}$ infers $v$ to a subset of candidates in $U$
(usually, the users in $U$ who have the most similar feature
distributions with that of $v$).
In this paper, we quantify the FDI for both approaches.
Specifically, in this subsection, we focus on distance-based FDI quantification.

To facilitate our quantification, we first make the following
definitions and assumptions.
For $x, y \in U \cup V$,
we define their feature distance as $D_{x, y}$.
In practice, $D_{x, y}$ can be defined in an application-oriented manner.
For instance, $D_{x, y}$ can be defined using the $\wp$-norm distance
as follows:
\begin{align*}
D_{x, y} = (\sum\limits_{i = 1}^N |g(f_x^i, w_x^i) - g(f_y^i, w_y^i)|^\wp)^{1/\wp}.
\end{align*}
Let $\mathrm{E}(\cdot)$ be the expectation/mean value
of a random variable.
Then, we define the expectation value of $D_{x, y}$
as $\mu_{x, y} = \mathrm{E}(D_{x, y})$.
Furthermore, we assume that
$D_{x, y} \in [0, \zeta_{x, y}]$, i.e., the feature distance
between $x$ and $y$ is lower bounded by 0 (which is an intuitive assumption) and
upper bounded by some value $\zeta_{x, y} \geq 0$.
Now, for $v \in V$ and $u, w \in U$, suppose that $v \simeq u$
and $v \neq w$.
We quantify the inferability of $v$ with respect to $u$ and $w$
in the following lemma.

\begin{lemma} \label{l3}
(1) When $\mu_{v, u} < \mu_{v, w}$, $v$ is inferable
if $\min\{\frac{1}{\zeta^2_{v, u}}, \frac{1}{2 \zeta^2_{v,w}}\}
\geq \frac{2(2 \ln N + 1)}{(\mu_{v, w} - \mu_{v, u})^2}$;
(2) When $\mu_{v, u} > \mu_{v, w}$, $v$ is inferable
if $\min\{\frac{1}{2\zeta^2_{v, u}}, \frac{1}{\zeta^2_{v, w}}\}
\geq \frac{2(2 \ln N + 1)}{(\mu_{v, u} - \mu_{v, w})^2}$.
\end{lemma}

{\em Proof}:
We start from proving the first conclusion.
Let $X = \frac{\mu_{v, u} + \mu_{v, w}}{2}$,
$\xi_1 = \frac{\mu_{v, w} - \mu_{v, u}}{2 \mu_{v, u}}$,
and $\xi_2 = \frac{\mu_{v, w} - \mu_{v, u}}{2 \mu_{v, w}}$.
When $\mu_{v, u} < \mu_{v, w}$, we have
\begin{align*}
& \Pr(D_{v, u} \geq D_{v, w}) \\
& \leq \Pr(D_{v, u} \geq X) + \Pr(D_{v, w} \leq X) \\
& = \Pr(D_{v, u} \geq (1 + \xi_1) \mu_{v, u}) + \Pr(D_{v, w} \leq (1 - \xi_2) \mu_{v, w}).
\end{align*}
Applying Chernoff bound (as shown in Lemma \ref{l_chernoff} in the Appendix),
we have
\begin{align*}
& \Pr(D_{v, u} \geq D_{v, w}) \\
& \leq \exp(- \frac{2 \xi_1^2 \mu_{v, u}^2}{\zeta^2_{v, u}})
+ \exp(-\frac{\xi_2^2 \mu_{v, w}^2}{\zeta^2_{v, w}}) \\
%& = \exp(- \frac{2 \mu_{v, u}^2}{\zeta^2_{v, u}} \cdot
%\frac{(\mu_{v, w} - \mu_{v, u})^2}{4 \mu_{v, u}^2}) +
%\exp(- \frac{\mu_{v, w}^2}{\zeta^2_{v, w}} \cdot
%\frac{(\mu_{v, w} - \mu_{v, u})^2}{4 \mu_{v, w}^2}) \\
& = \exp(- \frac{(\mu_{v, w} - \mu_{v, u})^2}{2 \zeta^2_{v, u}})
+ \exp(- \frac{(\mu_{v, w} - \mu_{v, u})^2}{4 \zeta^2_{v, w}}) \\
& \leq \max\{2 \exp(- \frac{(\mu_{v, w} - \mu_{v, u})^2}{2 \zeta^2_{v, u}}),
2 \exp(- \frac{(\mu_{v, w} - \mu_{v, u})^2}{4 \zeta^2_{v, w}})\} \\
& \leq 2\exp(-2 \ln N - 1)
 < 1/N^2.
\end{align*}
According to the Borel-Cantelli Lemma, we have
$\Pr(D_{v, u} \geq D_{v, w}) \stackrel{N \rightarrow \infty} = 0$ when $\mu_{v, u} < \mu_{v, w}$,
i.e., $\Pr(D_{v, u} < D_{v, w}) \stackrel{N \rightarrow \infty} = 1$.
Therefore, by comparing the feature distance, we can distinguish $v$
from $u$ and $w$, i.e., $v$ is inferable with respect to $u$ and $w$.

Now, we prove the second conclusion. When $\mu_{v, u} > \mu_{v, w}$,
Let $X = \frac{\mu_{v, u} + \mu_{v, w}}{2}$,
$\xi_1 = \frac{\mu_{v, u} - \mu_{v, w}}{2 \mu_{v, u}}$,
and $\xi_2 = \frac{\mu_{v, u} - \mu_{v, w}}{2 \mu_{v, w}}$.
Then, we have
\begin{align*}
& \Pr(D_{v, u} \leq D_{v, w}) \\
& \leq \Pr(D_{v, u} \leq X) + \Pr(D_{v, w} \geq X) \\
& = \Pr(D_{v, u} \leq (1 - \xi_1) \mu_{v, u}) + \Pr(D_{v, w} \geq (1 + \xi_2) \mu_{v, w}).
\end{align*}
Applying Chernoff bound,
we have
\begin{align*}
& \Pr(D_{v, u} \leq D_{v, w}) \\
& \leq \exp(- \frac{\xi_1^2 \mu_{v, u}^2}{\zeta^2_{v, u}})
+ \exp(-\frac{2 \xi_2^2 \mu_{v, w}^2}{\zeta^2_{v, w}}) \\
& = \exp(- \frac{\mu_{v, u}^2}{\zeta^2_{v, u}} \cdot
\frac{(\mu_{v, u} - \mu_{v, w})^2}{4 \mu_{v, u}^2}) \\
& \ \ \ \ \ +
\exp(- \frac{2 \mu_{v, w}^2}{\zeta^2_{v, w}} \cdot
\frac{(\mu_{v, u} - \mu_{v, w})^2}{4 \mu_{v, w}^2}) \\
& = \exp(- \frac{(\mu_{v, w} - \mu_{v, u})^2}{4 \zeta^2_{v, u}})
+ \exp(- \frac{(\mu_{v, w} - \mu_{v, u})^2}{2 \zeta^2_{v, w}}) \\
& \leq \max\{2 \exp(- \frac{(\mu_{v, w} - \mu_{v, u})^2}{4 \zeta^2_{v, u}}),
2 \exp(- \frac{(\mu_{v, w} - \mu_{v, u})^2}{2 \zeta^2_{v, w}})\} \\
& \leq 2\exp(-2 \ln N - 1)
 < 1/N^2.
\end{align*}
Thus, we have $\Pr(D_{v, u} \leq D_{v, w}) \stackrel{N \rightarrow \infty} = 0$,
i.e., $\Pr(D_{v, u} > D_{v, w}) \stackrel{N \rightarrow \infty} = 1$.
Therefore, $v$ is inferable with respect to $u$ and $w$
by comparing the feature distance.
\hfill $\Box$

In Lemma \ref{l3}, we quantify the feature distance-based FDI
conditions of $v$ with respect to $u$ and $w$.
In fact, the proof of Lemma \ref{l3} corresponds to
an implementation of $\mathcal{M}$: when the specified conditions
are satisfied, using a procedure as shown in Algorithm \ref{a_naive}
can make $v$ inferable with respect to $u$ and $w$ (now, we should
change $\Gamma_{\cdot \oplus \cdot}$ to $D_{\cdot, \cdot}$).
Also, $\mathcal{M}$ can be implemented using other techniques,
e.g., \cite{wanwansdm14}-\cite{narshmsp08}.
When the conditions are satisfied, as long as $\mathcal{M}$
is an increasing function on $D_{\cdot, \cdot}$,
$\mathcal{M}$ can successfully infer $v$ with respect
to $u$ and $w$.

Now, based on Lemma \ref{l3}, we study the Top-$K$
inferability of $v \in V$ with respect to $U$.
Again, we assume that $\exists u \in U$ such that $v \simeq u$.
The Top-$K$ FDI of $v$ is quantified in the following lemma.

\begin{lemma} \label{l4}
$v$ is Top-$K$ inferable if $\exists \overline{\mathcal{K}_v} \subseteq U$ such that
$|\overline{\mathcal{K}_v}| \geq n - K$,
$\mu_{v, u} \neq \mu_{v, x}$ for $\forall x \in \overline{\mathcal{K}_v}$,
and
$\min\{\frac{1}{\zeta_{v, x}^2} | x \in \{u\} \cup \overline{\mathcal{K}_v}\}
\geq \frac{8\ln N + 4\ln 2 \theta n}{\mu_{\min}}$,
where $\theta = \frac{|\overline{\mathcal{K}_v}|}{n}$ and $\mu_{\min} = \min\{(\mu_{v, u} - \mu_{v, x})^2 |
x \in \overline{\mathcal{K}_v}\}$.
\end{lemma}

{\em Proof}:
To prove this lemma, it is sufficient to prove that $\exists \mathcal{M}$
such that $\mathcal{M}(v : U) = \mathcal{K}_v$, $u \in \mathcal{K}_v$,
and $|\mathcal{K}_v| \leq K$.
Let $E$ be the event that $\exists w \in \overline{\mathcal{K}_v}$,
$v$ is not inferable with respect to $\{u, w\}$.
Then,
\begin{align*}
\Pr(E)
& = \Pr(\bigcup\limits_{w \in \overline{\mathcal{K}_v}}
v \text{ is not inferable with respect to } \{u, w\}) \\
& \leq \sum\limits_{w \in \overline{\mathcal{K}_v}}
\Pr(v \text{ is not inferable with respect to } \{u, w\}).
\end{align*}
Since $\mu_{v, u} \neq \mu_{v, x}$ for $\forall x \in \overline{\mathcal{K}_v}$
and based on Lemma \ref{l1}, we have
$\Pr(v$ is not inferable with respect to $\{u, w\})
\leq 2 \exp(-2 \ln N - \ln 2\theta n)$ (this can be proven
by considering $\mu_{v, u} > \mu_{v, w}$ and
$\mu_{v, u} < \mu_{v, w}$ respectively).
Therefore, we have
\begin{align*}
\Pr(E)
& \leq \sum\limits_{w \in \overline{\mathcal{K}_v}}
2 \exp(-2 \ln N - \ln 2\theta n) \\
& = 2 \theta n \exp(-2 \ln N - \ln 2\theta n)
 = 1/ N^2.
\end{align*}
Therefore, $\Pr(E) \stackrel{N \rightarrow \infty} = 0$,
which implies that
$\Pr(\forall w \in \overline{\mathcal{K}_v}, v$
is inferable with respect to $\{u, w\}) \stackrel{N \rightarrow \infty} = 1$.

Now, let $\mathcal{M}$ be the procedure as shown in Algorithm \ref{a_topnaive}
while changing $\Gamma_{v \oplus \cdot}$ to $D_{v \oplus \cdot}$.
Based on our proof, we conclude that the obtained $\mathcal{K}_v$ of Algorithm \ref{a_topnaive}
satisfies that $u \in \mathcal{K}_v$ and $\mathcal{K}_v \leq K$
(actually, $\mathcal{K}_v = K$).
\hfill $\Box$

In Lemma \ref{l4}, we quantified the conditions for a user $v \in V$
to be Top-$K$ inferable.
Based on Lemma \ref{l3} and Lemma \ref{l4},
we can quantify the $(\delta, K)$-inferability of $V$.
We show the result in the following theorem.

\begin{theorem} \label{t2}
Let $V_\delta$ be any subset of $\widetilde{V}$ with $|V_\delta| = \delta \widetilde{m}$.
$V$ is $(\delta, K)$-inferable if for $\forall v \in V_\delta$,
$\exists \overline{\mathcal{K}_v} \subseteq U$ such that
$|\overline{\mathcal{K}_v}| = n - K$,
$\mu_{v, u} \neq \mu_{v, x}$ for $\forall x \in \overline{\mathcal{K}_v}$,
and
$\min\{\frac{1}{\zeta_{v, x}^2} | x \in \{u\} \cup \overline{\mathcal{K}_v}\}
\geq \frac{8\ln N + 4\ln 2 \delta \theta \widetilde{m} n}{\mu_{\min}}$,
where $v \simeq u \in U$,
$\theta = \frac{n - K}{n}$ and $\mu_{\min} = \min\{(\mu_{v, u} - \mu_{v, x})^2 |
x \in \overline{\mathcal{K}_v}\}$.
\end{theorem}

{\em Proof}:
To prove this theorem, we take a similar approach
as in proving Theorem \ref{t1}.
Let $E$ be the event that $\exists v \in V_\delta$,
$v$ is not Top-$K$ inferable.
Then, we have
\begin{align*}
\Pr(E)
& = \Pr(\bigcup\limits_{v \in V_\delta} v \text{ is not Top}-K \text{ inferable}) \\
& \leq \sum\limits_{v \in V_\delta} \Pr(v \text{ is not Top}-K \text{ inferable}) \\
& \leq \sum\limits_{v \in V_\delta} \sum\limits_{w \in \overline{\mathcal{K}_v}}
2 \exp(-2\ln N - \ln 2 \delta \theta \widetilde{m} n) \\
& = \sum\limits_{v \in V_\delta} 2\theta n \exp(-2\ln N - \ln 2 \delta \theta \widetilde{m} n) \\
& = 2 \delta \theta \widetilde{m} n \cdot \exp(-2\ln N - \ln 2 \delta \theta \widetilde{m} n) \\
& = 1/N^2.
\end{align*}
Therefore, $\Pr(E) \stackrel{N \rightarrow \infty} = 0$,
which implies that
$\Pr(\forall v \in V_\delta, v$ is Top-$K$ inferable$) \stackrel{N \rightarrow \infty} = 1$,
i.e., $V$ is $(\delta, K)$-inferable.
\hfill $\Box$

In Theorem \ref{t2}, we quantify the feature distance-based
$(\delta, K)$-FDI of $V$.
When the specified conditions are satisfied, a $\mathcal{M}$ can be
constructed on top of the procedure in Algorithm \ref{a_topnaive}
(changing the $\Gamma_{x \oplus y}$-items to the $D_{x, y}$-items):
call Algorithm \ref{a_topnaive} for each user $v \in V$.
Then, according to the similar argument as in Lemma \ref{l2},
we can show that $V$ is $(\delta, K)$-inferable under $\mathcal{M}$.
Again, since the conditions quantified in Theorem \ref{t2}
(as well as in Lemma \ref{l3} and Lemma \ref{l4})
are sufficient while not necessary, it is possible to design
some sophisticated $\mathcal{M}$ to achieve better inference performance.

\subsection{General Quantification: From the Distribution Perspective}

In the previous subsection, we conduct the FDI quantification
for the applications that $\mathcal{M}$ employs a feature distance-based
inference model. In many other applications, $\mathcal{M}$ may employ a
feature distribution-based inference model \cite{wanwansdm14}-\cite{narshmsp08},
i.e., determine whether $v \in V$ and $u \in U$ are the same user
(or the data generated by the same user) according to the feature
distribution similarity of $v$ and $u$.
To provide the theoretical foundation for this kind of inference models,
we quantify the feature distribution-based FDI in this subsection.

For $v \in V$ and $u \in U$, there are many approaches to measure the
distribution similarity of
$\overrightarrow{\mathcal{F}(v)} = <g(f^i_v, w^i_v) | i = 1, 2, \cdots, N>$
and $\overrightarrow{\mathcal{F}(u)} = <g(f^i_u, w^i_u) | i = 1, 2, \cdots, N>$.
Among them, one of the most widely adopted approaches is the Cosine-similarity
based method \cite{davkulicml07}\cite{gershoccs14}\cite{baltrosp12}\cite{afrcalsp14}.
Therefore, we focus on quantifying the Cosine similarity-based FDI
in this paper. Our technique is expected to shed light on
the FDI quantification based on other distribution similarity measurements.
Before the quantification, we formally define the Cosine similarity first.
Let $x, y \in V \cup U$, $\overrightarrow{\mathcal{F}(x)} = <g(f^i_x, w^i_x) | i = 1, 2, \cdots, N>$,
and $\overrightarrow{\mathcal{F}(y)} = <g(f^i_y, w^i_y) | i = 1, 2, \cdots, N>$.
Furthermore, let $\mathrm{x} = ||\overrightarrow{\mathcal{F}(x)}||
= \sqrt{\sum\limits_{i = 1}^N (g(f^i_x, w^i_x))^2}$
be the \emph{magnitude} of a vector and $g_x^i = g(f_x^i, w_x^i)$.
Then, we define the feature distribution similarity between $x$ and $y$
as
\begin{align*}
\cos(x, y)
= \frac{\overrightarrow{\mathcal{F}(x)} \cdot \overrightarrow{\mathcal{F}(y)}}
{\mathrm{x}\mathrm{y}}
= \frac{\sum_{i = 1}^N g_x^i \times g_y^i}
{\mathrm{x}\times \mathrm{y}},
\end{align*}
where the $\cdot$ is the \emph{dot product} here.

Now, given $v \in V$ and $u, w \in U$, we assume that $v \simeq u$ and $v \neq w$.
We start our quantification from the scenario
that $v$ is inferable with respect to $\{u, w\}$.
Let $X_i$ and $X$ be two random variables such that
$X_i = g_v^i (\mathrm{w} g_u^i - \mathrm{u} g_w^i)$
and $X = \sum\limits_{i = 1}^N X_i$.
Furthermore, we assume that $X_i \in [l, h]$.
Then, we have the following lemma to quantify the inferability of $v$
with respect to $\{u, w\}$.

\begin{lemma} \label{l5}
$v$ is inferable with respect to $\{u, w\}$
if $\mu \geq \frac{(h - l) \sqrt{2N \ln N}}{\xi}$,
where $\mu = \mathrm{E}(X)$ is the expectation value of $X$
and $\xi \in (0, 1)$ is a constant value.
\end{lemma}

{\em Proof}:
To prove this lemma, statistically, it is sufficient to prove that
$\Pr(\cos(v, u) > \cos(v, w)) = \Pr(\cos(v, u) - \cos(v, w) > 0)
\rightarrow 1$ as $N \rightarrow \infty$.
According to the Cosine similarity definition, we have
\begin{align*}
\cos(v, u) - \cos(v, w)
& = \frac{\sum_{i = 1}^N g_v^i  g_u^i} {\mathrm{v} \mathrm{u}}
- \frac{\sum_{i = 1}^N g_v^i  g_w^i} {\mathrm{v} \mathrm{w}} \\
& = \frac{\mathrm{w} \sum_{i = 1}^N g_v^i  g_u^i
- \mathrm{u} \sum_{i = 1}^N g_v^i  g_w^i }
{\mathrm{v} \mathrm{u} \mathrm{w}} \\
& = \frac{\sum_{i = 1}^N g_v^i (\mathrm{w} g_u^i - \mathrm{u} g_w^i)}
{\mathrm{v} \mathrm{u} \mathrm{w}} \\
& = \frac{X} {\mathrm{v} \mathrm{u} \mathrm{w}}.
\end{align*}
Therefore, to prove $\Pr(\cos(v, u) - \cos(v, w) > 0)
\stackrel{N \rightarrow \infty} {\rightarrow} 1$, it is equivalent to prove that
$\Pr(X > 0) \stackrel{N \rightarrow \infty} {\rightarrow} 1$.
Now, instead of proving $\Pr(X > 0) \stackrel{N \rightarrow \infty} {\rightarrow} 1$
directly, we prove
$\Pr(X \leq \epsilon) \stackrel{N \rightarrow \infty} {\rightarrow} 0$,
where $\epsilon \in (0, (1 - \xi) \mu)$ is some constant value.
According to the Chernoff bound, we have
\begin{align*}
\Pr(X \leq \epsilon)
& \leq \Pr(X \leq (1 - \xi) \mu) \\
& \leq \exp(-\frac{\xi^2 \mu^2}{N(h - l)^2}) \\
& \leq 1/N^2.
\end{align*}
Thus, $\Pr(X \leq \epsilon) \stackrel{N \rightarrow \infty} {\rightarrow} 0$,
i.e., $\Pr(\cos(v, u) - \cos(v, w) > 0)
\stackrel{N \rightarrow \infty} {\rightarrow} 1$,
which implies $v$ is inferable with respect to $\{u,  w\}$.
\hfill $\Box$

In Lemma \ref{l5}, we quantify the inferability of $v$ with respect to $\{u, w\}$.
Following the proof of the lemma, a $\mathcal{M}$ can be easily constructed
such that $\mathcal{M}(v, \{u, w\}) = \{u\}$ when the specified conditions
are satisfied: $\mathcal{M}$ simply returns the one who has
a higher feature distribution similarity with $v$.
Based on Lemma \ref{l5}, we can further quantify the Top-$K$
inferability of $v$.
The result is shown in Lemma \ref{l6}.

\begin{lemma} \label{l6}
$v$ is Top-$K$ inferable
if $\mu \geq \frac{(h - l) \sqrt{N (\ln N^2 \theta n)}}{\xi}$,
where $\mu = \mathrm{E}(X)$, $\theta = \frac{n-K}{n}$,
and $\xi \in (0, 1)$ is a constant value.
\end{lemma}

{\em Proof}:
This lemma can be proven based on Lemma \ref{l5}.
Let $\overline{\mathcal{K}_v}$ be any subset of $U$ such that
$|\overline{\mathcal{K}_v}| = n - K$ and $u \notin  \overline{\mathcal{K}_v}$.
Then, we first prove that $\Pr(\forall w \in \overline{\mathcal{K}_v},\cos(v, u) - \cos(v, w) > 0)
\stackrel{N \rightarrow \infty} {\rightarrow} 1$.
Let $E$ be the event that $\exists w \in \overline{\mathcal{K}_v}$
such that $\cos(v, u) \leq \cos(v, w)$.
Then, applying Lemma \ref{l5}, we have
\begin{align*}
\Pr(E)
& = \Pr(\bigcup\limits_{w \in \overline{\mathcal{K}_v}}
\cos(v, u) \leq \cos(v, w)) \\
& \leq \sum \limits_{w \in \overline{\mathcal{K}_v}}
\Pr(\cos(v, u) \leq \cos(v, w)) \\
& \leq \sum \limits_{w \in \overline{\mathcal{K}_v}} \exp(-\frac{\xi^2 \mu^2}{N(h - l)^2}) \\
& \leq \sum \limits_{w \in \overline{\mathcal{K}_v}} \exp(- \ln \theta n N^2) \\
& = \theta n \exp(- \ln \theta n N^2)
 = 1/N^2.
\end{align*}
Therefore, we have $\Pr(\forall w \in \overline{\mathcal{K}_v},\cos(v, u) - \cos(v, w) > 0)
\stackrel{N \rightarrow \infty} {\rightarrow} 1$.
Now, we design a $\mathcal{M}$ for $v$ to be Top-$K$ inferable.
Similar to the one in Algorithm \ref{a_topnaive},
we can design a $\mathcal{M}$ under which
the users in $U$ who have the Top-$K$ feature distribution similarity scores
(Cosine similarity scores) with $v$ are returned as $\mathcal{K}_v$.
Then, based on our proof, we have $u \in \mathcal{K}_v$.
\hfill $\Box$

In Lemma \ref{l6}, the feature distribution-based
Top-$K$ inferability of $v$ is quantified.
When the specified conditions are satisfied,
we also discussed how to implement $\mathcal{M}$
in the proof. Based on Lemma \ref{l5} and Lemma \ref{l6}
we can quantify the $(\delta, K)$-inferability of $V$.
The result is shown in the following theorem.

\begin{theorem} \label{t3}
$V$ is $(\delta, K)$-inferable
if $\mu \geq \frac{(h - l) \sqrt{N \ln (\delta \theta \widetilde{m} n N^2)}}{\xi}$,
where $\theta = \frac{n-K}{n}$ and
and $\xi \in (0, 1)$ is a constant value.
\end{theorem}

{\em Proof}:
Let $V_\delta$ be any subset of $\widetilde{V}$ with size $\delta \widetilde{m}$.
To prove this theorem, it is sufficient to prove that
all the users in $V_\delta$ are Top-$K$ inferable.
Let $E$ be the event that $\exists v \in V_\delta$ such that
$v$ cannot be Top-$K$ inferable. Then, we have
\begin{align*}
\Pr(E)
& = \Pr(\bigcup\limits_{v \in V_\delta} v \text{ is not Top-}K \text{ inferable}) \\
& \leq \sum\limits_{v \in V_\delta} \Pr(v \text{ is not Top-}K \text{ inferable}).
\end{align*}
According to Lemma \ref{l6}, we have
\begin{align*}
\Pr(E)
& \leq \sum\limits_{v \in V_\delta} \sum \limits_{w \in \overline{\mathcal{K}_v}}
\exp(-\frac{\xi^2 \mu^2}{N(h - l)^2}) \\
& \leq \sum\limits_{v \in V_\delta} \sum \limits_{w \in \overline{\mathcal{K}_v}}
\exp(- \ln (\delta \theta \widetilde{m} N^3)) \\
& = \delta \widetilde{m} \cdot \theta n \cdot \exp(- \ln (\delta \theta \widetilde{m} n N^2)) \\
& = 1/N^2.
\end{align*}
Therefore, we have $\Pr(E)\stackrel{N \rightarrow \infty} {\rightarrow} 0$,
i.e., statistically, all the users in $V_\delta$ are $(\delta, K)$-inferable.
\hfill $\Box$

In Theorem \ref{t3}, we quantify the feature distribution similarity-based
$(\delta, K)$-FDI of $V$. When the specified conditions are satisfied,
we can also design a $\mathcal{M}$ using the one shown in Lemma \ref{l6}:
finding the $\mathcal{K}_v$ for each user $v \in V$ using the $\mathcal{M}$
shown in Lemma \ref{l6}. According to our proof,
we can see that $V$ is $(\delta, K)$-inferable under such a $\mathcal{M}$.
Furthermore, similar to that in Theorem \ref{t1} and Theorem \ref{t2},
the conditions in Theorem \ref{t3} are sufficient while not necessary.
Therefore, a sophisticated $\mathcal{M}$ could be implemented to improve
the inference performance. Here, our FDI quantification can serve as a
theoretical baseline to facilitate and guide the design of better
inference models.

\subsection{Discussion: Inferring New User/Data}

In the previous subsections, we focus on quantifying the
feature distance and distribution based FDI of the users
that appear in both the training data $U$ and the target data $V$.
In reality, it is possible that there are some new users/data
that appear in $V$ while not in $U$.
Formally, it is possible that $\exists v' \in V$ while $\nexists u \in U$
such that $v' \simeq u$.
In this case, an ideal inference model $\mathcal{M}$ will
infer $v'$ as a new user (or data generated by a new user),
e.g., an intruder in network forensics applications \cite{wanwansdm14}\cite{gershoccs14}.
In practical inference models \cite{wanwansdm14}-\cite{narshmsp08},
a user $v'$ in $V$ is inferred as a new user (or data generated
by a new user) if the feature distance $D_{v', u}$ is larger than a threshold
for $\forall u \in U$,
or the feature distribution similarity $\cos(v', u)$
is smaller than a threshold for $\forall u \in U$.

Theoretically, it is challengeable (or, impossible) to quantify the precise inferability
of a new user $v' \in V$ in general with statistical guarantee
(that is why an inference system has false positive and false negative).
The reason is that theoretically, the feature characteristics
of a new user (data generated by a new user) might be arbitrarily
similar to an existing user (e.g., the network intruders
keep improving their camouflaging techniques).
Nevertheless, our FDI quantification still has meaning implications
for inferring new users.
For $x \in V$, $y \in U$, and $x \simeq y$,
let $\mu^*_d = \mathrm{E}(D_{x, y})$
when $\mathcal{M}$ is a feature distance based model
and $\mu^*_s = \mathrm{E}(\cos(x, y))$ when $\mathcal{M}$ is a feature
distribution similarity based model.
Then, when $D_{v', u}$ is significantly apart from
$\mu^*_d$ or $\mu_s^*$ depending on $\mathcal{M}$
(distance or distribution based),
$v'$ can be inferred as a new user (the data generated by a new user)
with a higher confidence, i.e., $\mu^*_d$ or $\mu_s^*$ can be set as
the threshold values in practical applications.
The behind-the-scene reason for this fact can be explained by
the following corollary, which is a direct result of the Chernoff bound.

\begin{corollary} \label{c3}
(1) Let $D_{\min} = \min \{D_{v', x} | v' \in V, x \in U\}$
and $D_{\max} = \max \{D_{v', x} | v' \in V, x \in U\}$.
When $\mu^*_d \in [0, \zeta]$ and $\mu^*_d \geq \frac{\zeta \sqrt{2\ln N}}{\xi}$,
$v'$ is a new user if $D_{\min} \geq (1 + \xi) \mu^*_d$
or $D_{\max} \leq (1 - \xi) \mu^*_d$ for all $\xi > 0$.
(2) Let $C_{\max} = \max \{\cos(v', x) | v' \in V, x \in U\}$.
When $\mu^*_s \in [l, h]$ and $\mu^*_s \geq \frac{(h - l) \sqrt{2\ln N}}{\xi}$,
$v'$ is a new user if $C_{\max} \leq (1 - \xi) \mu^*_s$  for all $\xi > 0$.
\end{corollary}

In practice, the accurate value of $\mu^*_d$ or $\mu_s^*$ is usually difficult
to be obtained, if not impossible. Frequently, $\mu^*_d$ or $\mu_s^*$ can only
be estimated based on the observed data and thus it may change with more data coming,
i.e., the threshold estimation problem itself is an interesting problem.
For our purpose, we propose to quantify the correlation between
the threshold setting and the false positive/negative rate of $\mathcal{M}$
as one of our future research directions.

\section{Evaluation} \label{evaluation}

In this section, we evaluate the user inferability of real world security and
privacy applications based on our FDI quantification.
Specifically, we evaluate two scenarios:
network traffic attribution in network forensics
and feature-based data de-anonymization (as shown in Section \ref{example}).

%In this section,
%we conduct evaluation on the user inferability
%of four large-scale network traces collected from a large enterprise in 2014
%based on our FDI quantification.

\subsection{Network Traffic Attribution}

\subsubsection{Data Collection and Analysis}

In this scenario, we evaluate the user inferability of four large-scale network
traces generated by the employees of a large enterprise.
These four traces are collected in four periods of 2014: April 1 -- April 30
which consists of the network traffic generated by 5888 users,
July 1 -- July 31 which consists of the network traffic generated by 5610 users,
October 1 -- October 31 which consists of the network traffic generated by 5268 users,
and December 1 -- December 31 which consists of the network traffic generated by 5699 users.
For each network trace, it is composed of three parts:
HTTP request headers, netflow measures, and DNS queries.

Here,
we do not consider the network traffic payloads, e.g., the HTTP payloads,
for the following reasons. First, those data are highly sensitive
and using them may cause some legal issues.
Second, although network traffic payloads may provide more information,
using our network traces is sufficient to infer many users as shown in
our experiments. Finally, as indicated in \cite{wanwansdm14},
in most of the common available traces,
they do not have those payloads. Therefore,
studying the common feature-based data inferability would be
more useful and general for security and privacy applications.

\begin{center}
\begin{table}[t!]
%\fontsize{7pt}{8pt}\selectfont
 \caption{Network trace features.} \label{tab_data}
  \centering
  \begin{tabular}{  c c c c}
    \hline
     & $n$ & $N$ & \# of user-feature relationships \\
     \hline
     Apr-Domain & 5,888 & 290,537 & 3,968,361 \\
     Apr-Path & 5,888 & 1,685,439 & 17,389,051 \\
     July-Domain & 5,610 & 391,290 & 3,739,246 \\
     July-Path & 5,610 & 1,855,415 & 16,010,442 \\
     Oct-Domain & 5,268 & 270,604 & 3,868,538 \\
     Oct-Path & 5,268 & 1,741,781 & 16,895,932 \\
     Dec-Domain & 5,699 & 298,490 & 3,736,956 \\
     Dec-Path & 5,699 & 2,159,448 & 16,926,145 \\
    \hline
  \end{tabular}
  %\vspace{-6mm}
  \end{table}
 % \vspace{-4mm}
\end{center}

\subsubsection{Feature Extraction}
After collecting these four traces, we extract the features of them.
Here, we use the feature extraction model proposed in
\cite{wanwansdm14}. Although we may extract more features,
for our purpose, it is sufficient to extract two kinds of lexical-based features
for our FDI analysis: \emph{domain feature} and \emph{path feature (tokenized)}.
Basically, these two features characterize the behaviors of users
in terms of \emph{what} types of websites they have visited
and \emph{how} they interacted with the websites.
For instance, given a HTTP request
``www.google.com/search?q=ndss+2016\&ie=utf-8\&oe=utf-8",
we will extract a domain feature as ``www.google.com".
For the path features, we tokenize each path (URL) using `?', `=',
`\&", etc. as delimiters and employ a bag-of-word representation of the tokens.
We refer to the interested readers for more details
of the feature extraction model to \cite{wanwansdm14}.
Finally, we show the feature extraction results of the
four traces in Table \ref{tab_data},
where Apr, July, Oct, and Dec represent the four traces collected in
April, July, October, and December of 2014,
``-Domain" means the domain features,
``-Path" means the tokenized path features,
$n$ is the number of users in the dataset,
and $N$ is the number extracted features.
Note that, for each user-feature relationship in Table \ref{tab_data},
there is a weight associated with it, which indicates how many times
that a feature appeared in a user's trace.
For instance, if Bob visited ``www.google.com" 100 times in April, 2014,
then the weight associated with the ``Bob -- www.google.com/" relationship
is 100 in the Apr-Domain dataset in Table \ref{tab_data}.

\begin{figure*}[!ht]
 \centering
  \subfigure[Domain]{
    \includegraphics[width=3in]{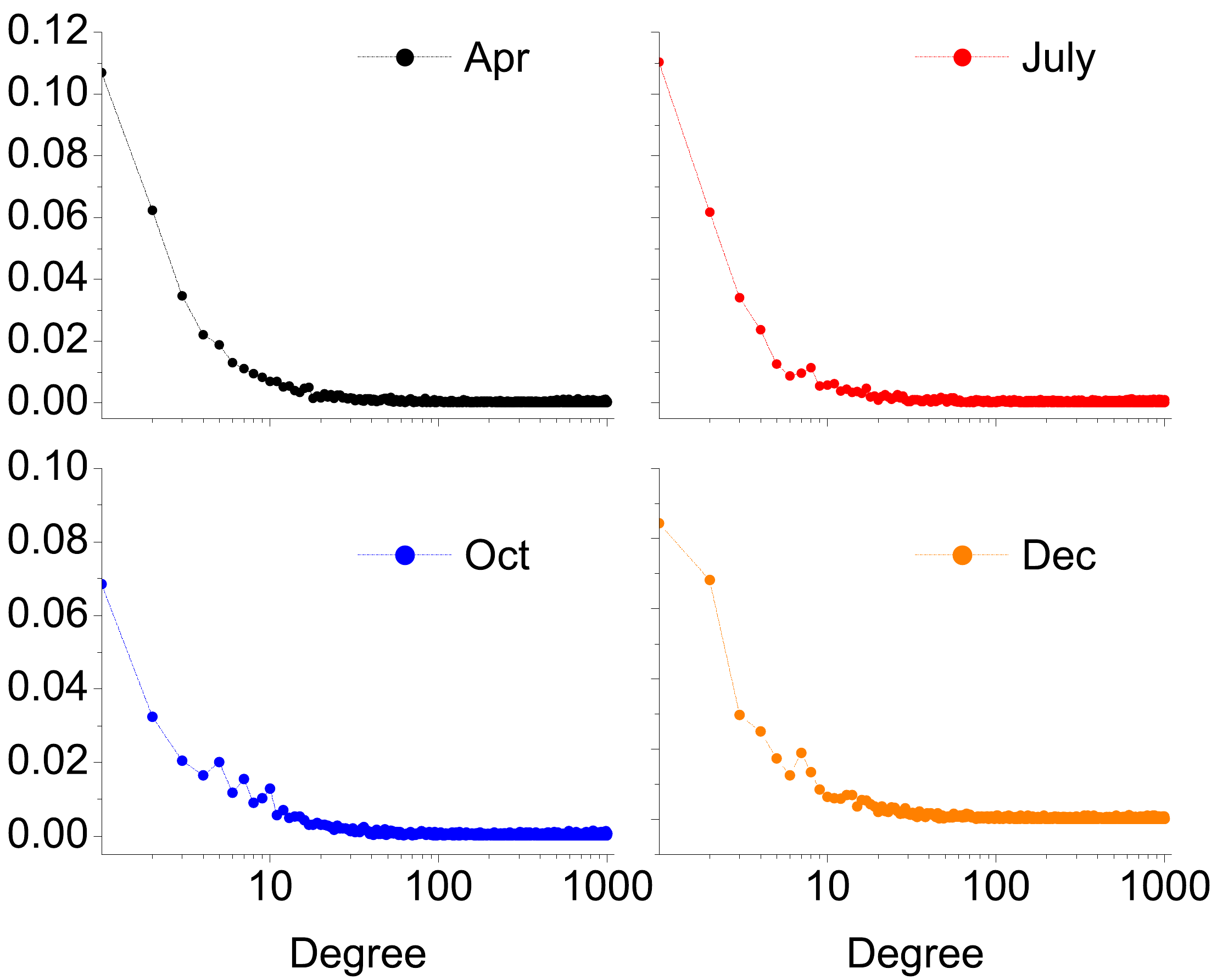}
  }
  \subfigure[Path]{
    \includegraphics[width=3in]{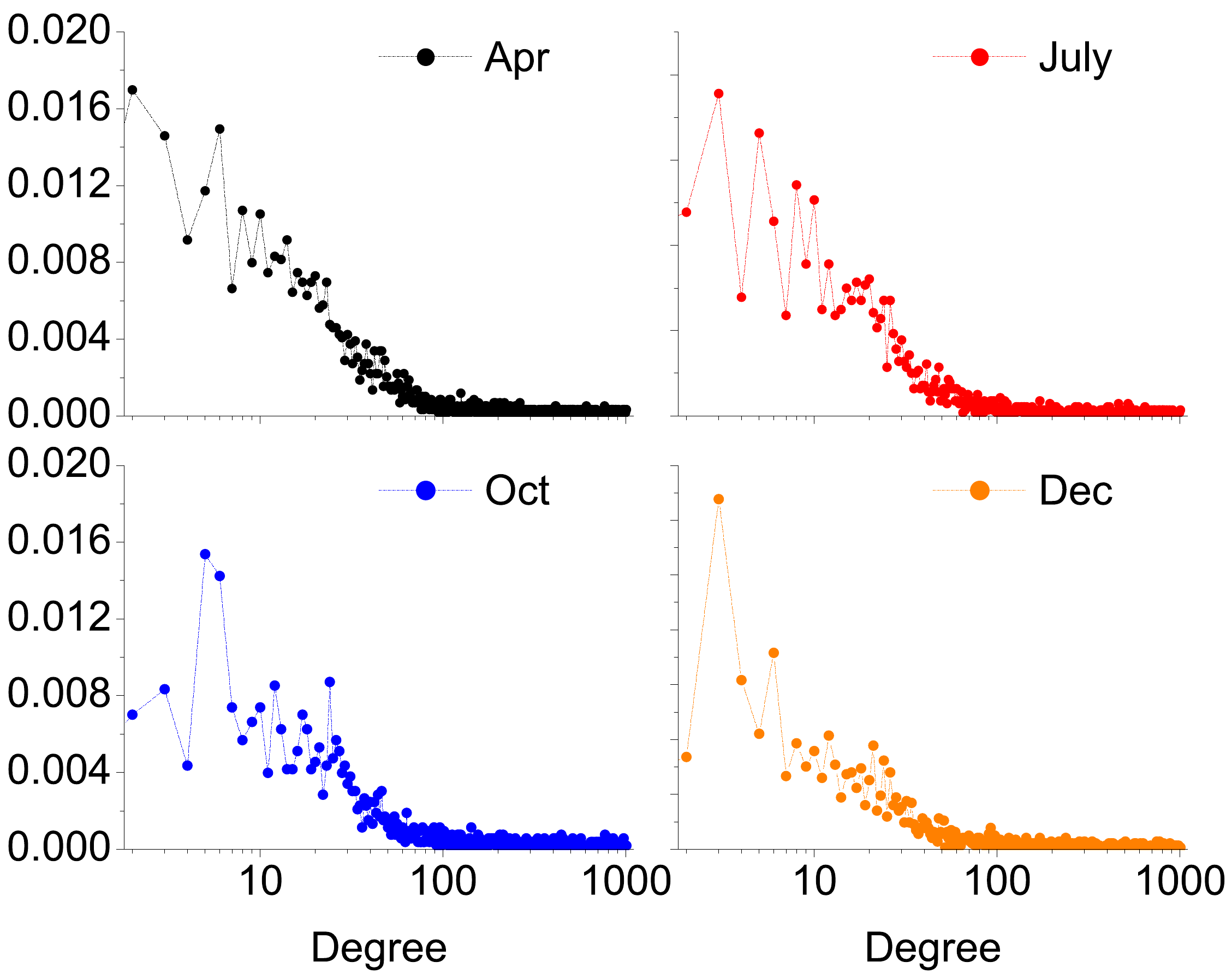}
  }
  \caption{User degree distribution.} \label{f_userdis}
%\vspace{-7mm}
\end{figure*}

\begin{figure*}[!ht]
 \centering
  \subfigure[Domain]{
    \includegraphics[width=3in]{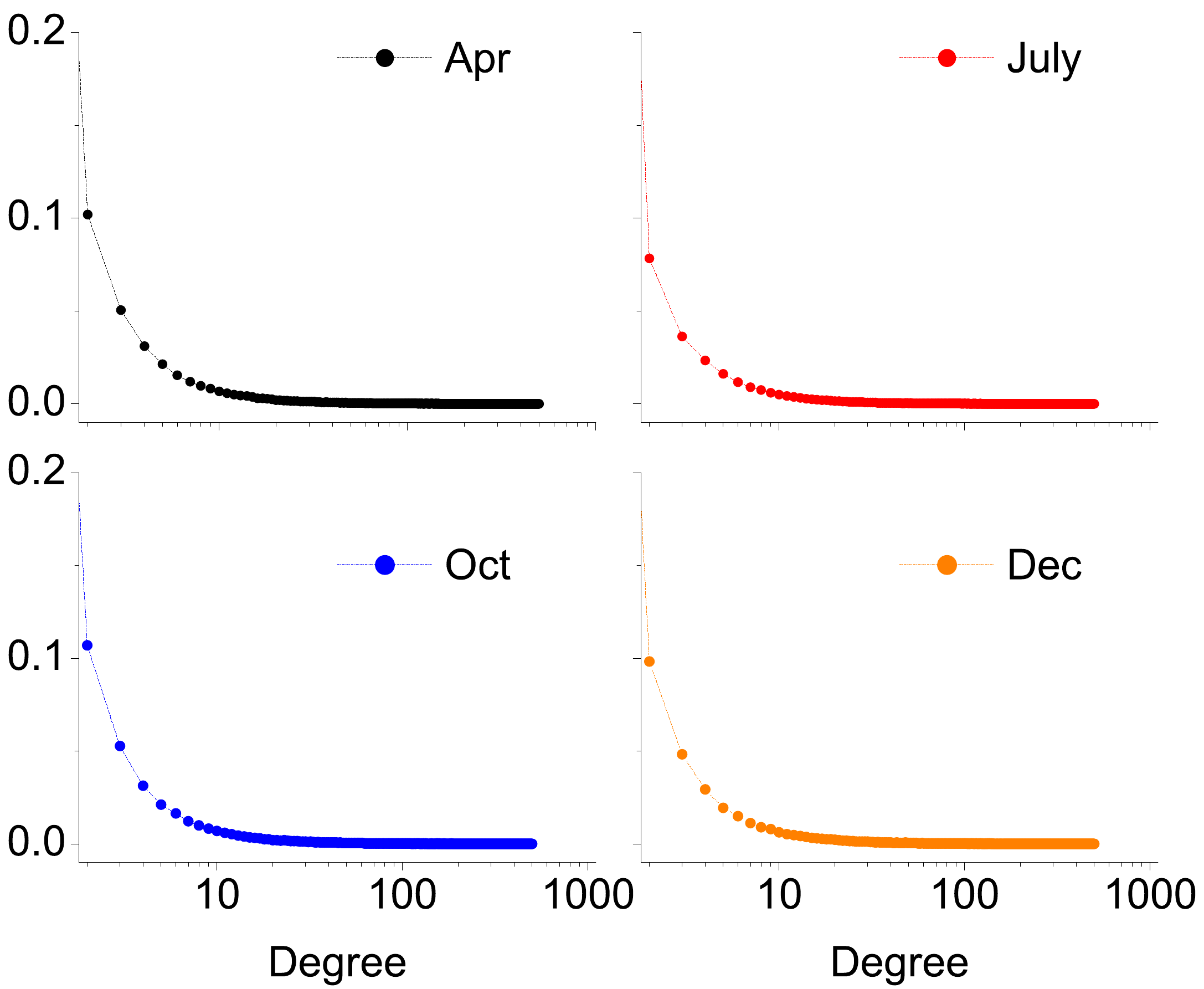}
  }
  \subfigure[Path]{
    \includegraphics[width=3in]{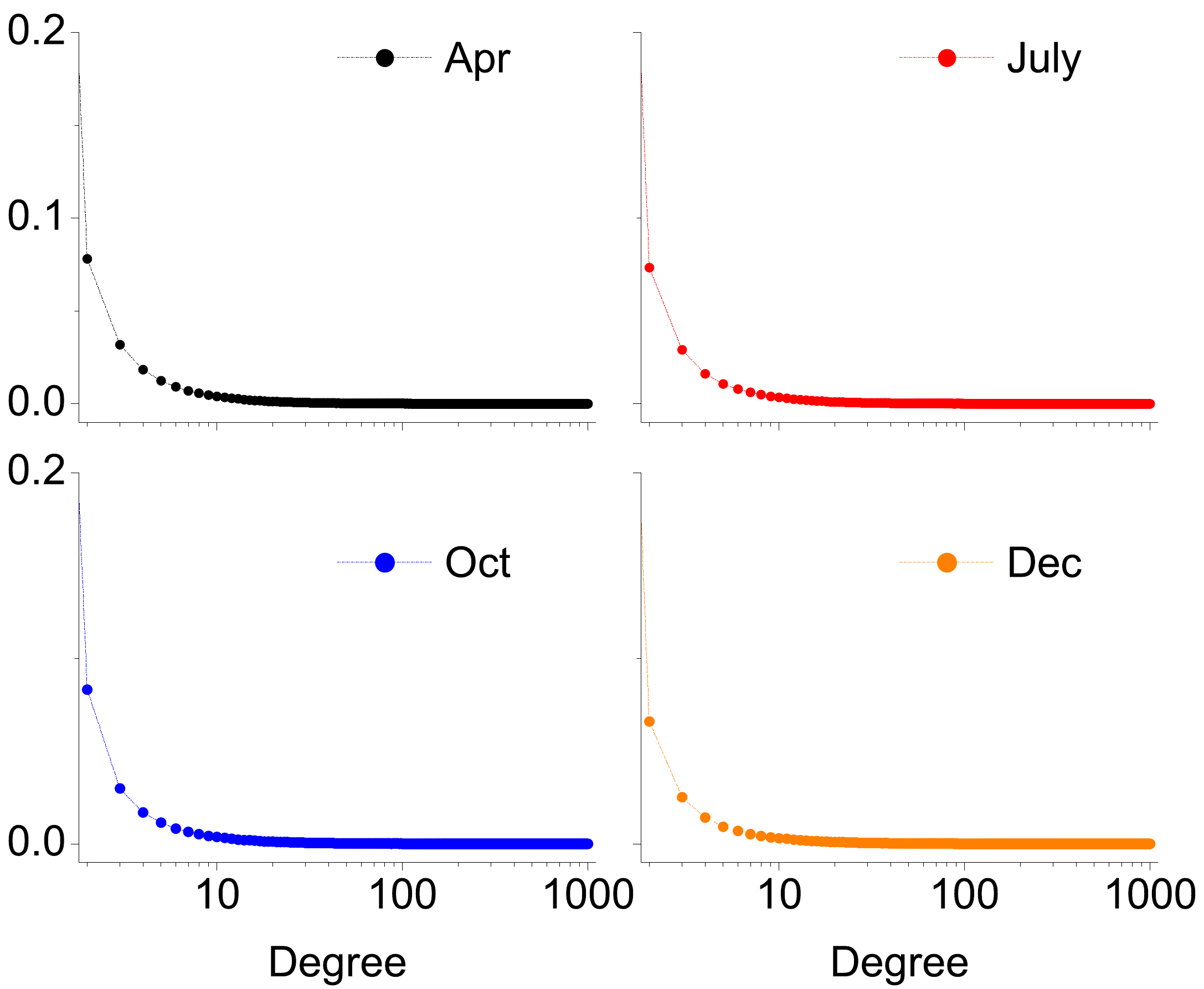}
  }
  \caption{Feature degree distribution.} \label{f_feadis}
%\vspace{-7mm}
\end{figure*}

Now, we define the \emph{degree} of each user
as the number of features this user has and the \emph{degree}
of each feature as the number users that have this feature.
Then, we show the user degree distribution and feature degree
distribution of the traces in Table \ref{tab_data} with respect to the domain feature and
the path feature in Fig.\ref{f_userdis} and Fig.\ref{f_feadis}, respectively.
From Fig.\ref{f_userdis} and Fig.\ref{f_feadis},
we have the following observations:
both the user degree and feature degree generally follow
a power-law-like distribution \cite{powerlaw}, especially the feature degree distribution,
i.e., most of the users have a small number of features while only a few users
have many features, and meantime,
most of the features only appear in the trace of a few users
while a small number of features appear in the trace of a large number of users.
These distributions together suggest that these features could
be employed to effectively infer the users.

\subsubsection{Evaluation Methodology}

To conduct the FDI evaluation, we basically follow the same process
as shown in Fig.\ref{f_example} and Fig.\ref{f_model}.
Meanwhile, since we focus on evaluating the statistically inherent
FDI, we also make the evaluation process mathematically tractable.
Following the models shown in Fig.\ref{f_example} and Fig.\ref{f_model},
we first determine the training data and the testing data.
Here, instead of partitioning the raw data into two parts for training
and testing respectively (as in many existing literature, e.g., \cite{wanwansdm14}),
we take another while theoretically equivalent approach:
following the FDI quantification in Section \ref{quantification},
we first construct the training dataset by keep all the users and features
in each trace
while sample the user-feature relationships independently and identically
using a probability $p$ \footnote{It is not necessary to have the training data
and the testing data to have the same group of users or features.
If they are not the same, we can either apply our theory to the overlapped users/features,
or make them the same by adding isolated users/features that only appeared in
the other dataset. Theoretically, different user/feature group will
not change the validity of our quantification.}.
Similarly, we construct the testing/targeting dataset using the same process
as in obtaining the training dataset.
We use this approach to construct the training and testing data
for two reasons. First, mathematically, this approach is equivalent
to the traditional method in \cite{wanwansdm14}.
In the traditional method, the raw data is partitioned into the training data
and the testing data and then features are extracted from both datasets.
Apparently, the reason that existing inferring techniques can work is that
the training data and the testing data share some common features
(or similar distributions over a feature space).
Therefore, statistically, we can consider the training and testing
data as some sampling versions of the original raw data respectively,
i.e., each training and testing data partition method
mathematically corresponding to one $p$ here.
Second, using this approach to obtain the training and testing/targeting data
makes it easier to apply our FDI quantification analysis.
We will make more discussions on closing the gap between theory and
practice in Section \ref{discussion}.

After obtaining the training and testing data,
we quantify the FDI of the four traces using the general scenario FDI
quantification technique in Section \ref{quantification}.
Specifically, for the network traffic attribution application,
most the of existing inference models are based on feature distance
\cite{wanwansdm14}\cite{davkulicml07}\cite{neaperccs14}.
Therefore, we evaluate the FDI using the distance-based quantification technique here.
Following Theorem \ref{t2}, we can easily construct an
inference model $\mathcal{M}$ on top of the procedure of Algorithm \ref{a_topnaive}
as shown in Section \ref{dispers}.
Then, we apply $\mathcal{M}$ to quantify the FDI of each dataset.
We also make more discussion on the implications of our quantification
as well as the implications of the results in Section \ref{discussion}.

\subsubsection{Results and Analysis}

\begin{figure*}[!ht]
 \centering
  \subfigure[Domain]{
    \includegraphics[width=3.3in]{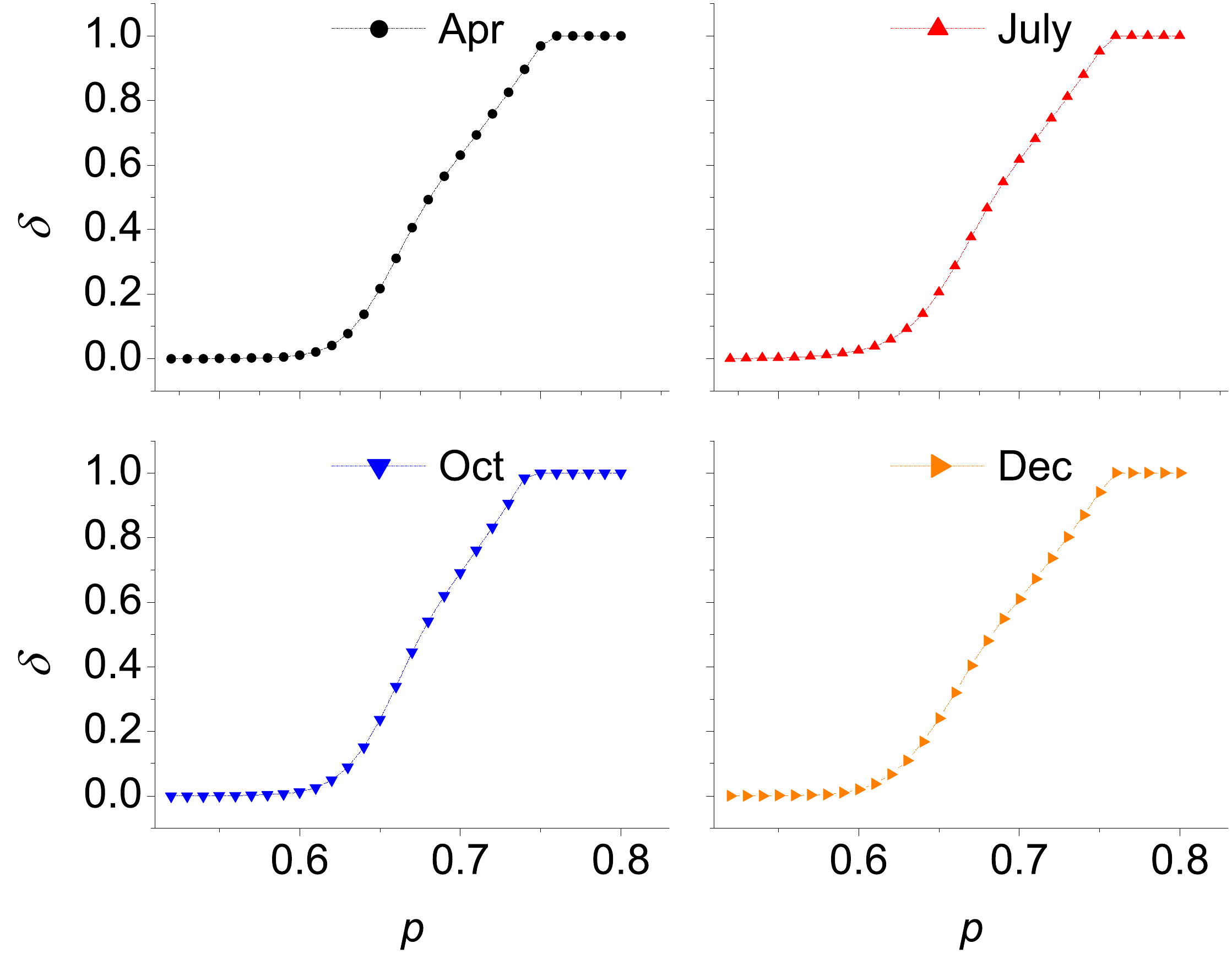}
  }
  \subfigure[Path]{
    \includegraphics[width=3.3in]{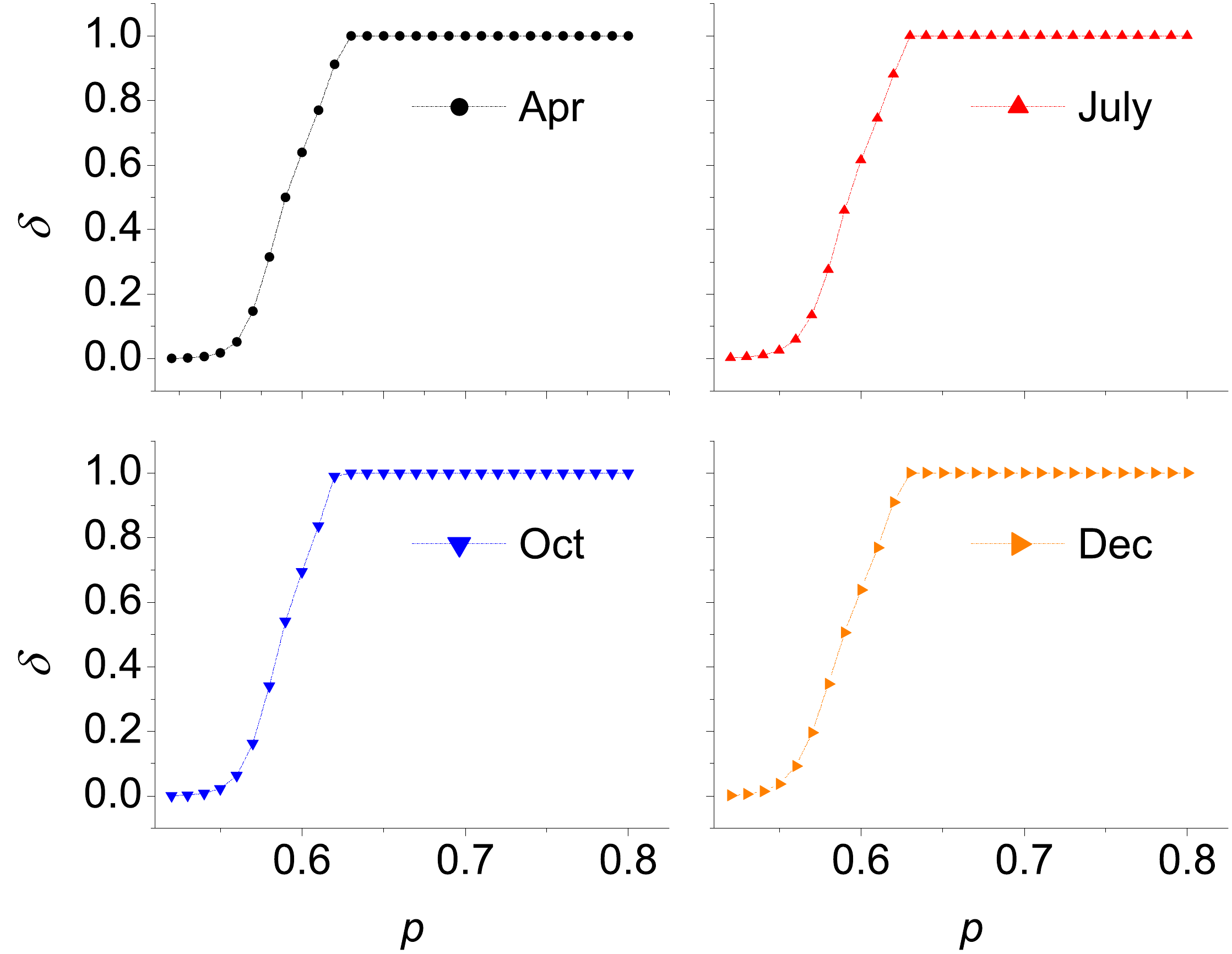}
  }
  \caption{$(\delta, K)$-inferable verses $p$.
  $K = 10$ in the experiment.} \label{f_ntd}
%\vspace{-7mm}
\end{figure*}

Now, we evaluate the FDI of the four traces following the above
evaluation methodology. To reduce any bias, all the experiments are run 10
times (e.g., for the same $p$). The final results are the
average of that of the 10 runs.
We show the $(\delta, K)$-inferability
of the four datasets with respect to the domain and path features in Fig.\ref{f_ntd}
respectively,
where we set $K = 10$, i.e., we are targeting a user to be
Top-10 inferable.
From Fig.\ref{f_ntd}, we have the following observations.
\begin{itemize}
\item
With the increase of $p$, $\delta$ also increases,
which implies that more and more users become Top-10 inferable.
The reason is that a large $p$ implies more common features
are shared by the training data and the targeting data,
i.e., there is more knowledge available to an inference model.
Therefore, statistically, more users can be successfully Top-10 inferable.

\item
When comparing the domain feature-based data inferability (Fig.\ref{f_ntd} (a))
with the path feature-based data inferability (Fig.\ref{f_ntd} (b)),
we find that the path features are more powerful in inferring
the users than the domain features.
This can be explained based on the results in Table \ref{tab_data},
Fig.\ref{f_userdis}, and Fig.\ref{f_feadis}.
First, for each dataset, it has much more path features than
domain features (Table \ref{tab_data}),
i.e., much more knowledge can be used to conduct path feature-based inference.
Second, the users of each dataset have higher path feature-based degrees
than domain feature-based degrees (Fig.\ref{f_userdis}),
and meanwhile, both the domain and the path feature degree distributions
generally follow similar power-law-like distributions.
Thus, users are more distinguishable with respect to the
path features than that of the domain features.
\end{itemize}

In our evaluation, we also examined the data inferability with respect to
other settings: changing the value of $K$ and
combining the domain and path features.
The results are as expected and we put them in the technical report \cite{report}.
Here, we briefly summarize the results.
When increasing $K$ (from $0.05n$ to $0.2n$), more users are Top-$K$ inferable
given the same $p$. The reason is evident since increasing $K$ implies
decreasing the desired inference accuracy.
Statistically, more users become Top-$K$ inferable.
Furthermore, after combining the domain and path features together,
we also have more users inferable compared to the scenario of applying the
domain and path features separately.
The reason is also straightforward since more features imply more knowledge
are available for inferring users, and thus the inference accuracy is improved.

\subsection{Data De-anonymization}

Now, we evaluate the users' feature-based inferability in the data de-anonymization
application \cite{calharusenix15}\cite{afrcalsp14}\cite{narshmsp08}.

\subsubsection{Data Collection and Features}

In our evaluation, we use three social network datasets,
Google+, Facebook, and Twitter, which are publicly available
at the Stanford Large Network Dataset Collection \cite{snap}.
The reason for us to use these datasets is that
they are published along with well-defined user features,
e.g., birthdays, education, hometown, languages, career, etc.
For de-anonymization attacks, an adversary may directly employ these features
to de-anonymize users. For our purpose, we can also employ these features
to quantify users' FDI.
We show the statistics of these three datasets
in Table \ref{tab_social}.
By comparing the datasets in Tables \ref{tab_data} and \ref{tab_social},
we can find that the three social datasets have much less features.
Furthermore, for the three datasets in \ref{tab_social},
there is no weight information associated with the user-feature relationships.

\begin{center}
\begin{table}[t!]
%\fontsize{7pt}{8pt}\selectfont
 \caption{Data statistics.} \label{tab_social}
  \centering
  \begin{tabular}{  c c c c}
    \hline
     & $n$ & $N$ & \# of user-feature relationships \\
     \hline
     Google+ & 107,614 & 19,044 & 387,261 \\
     Facebook & 4,039 & 1,283 & 37,257 \\
     Twiiter & 81,306 & 216,839 & 1,245,234 \\
    \hline
  \end{tabular}
  %\vspace{-1mm}
  \end{table}
% \vspace{-4mm}
\end{center}

\begin{figure*}[!ht]
 \centering
  \subfigure[Google+]{
    \includegraphics[width=1.8in]{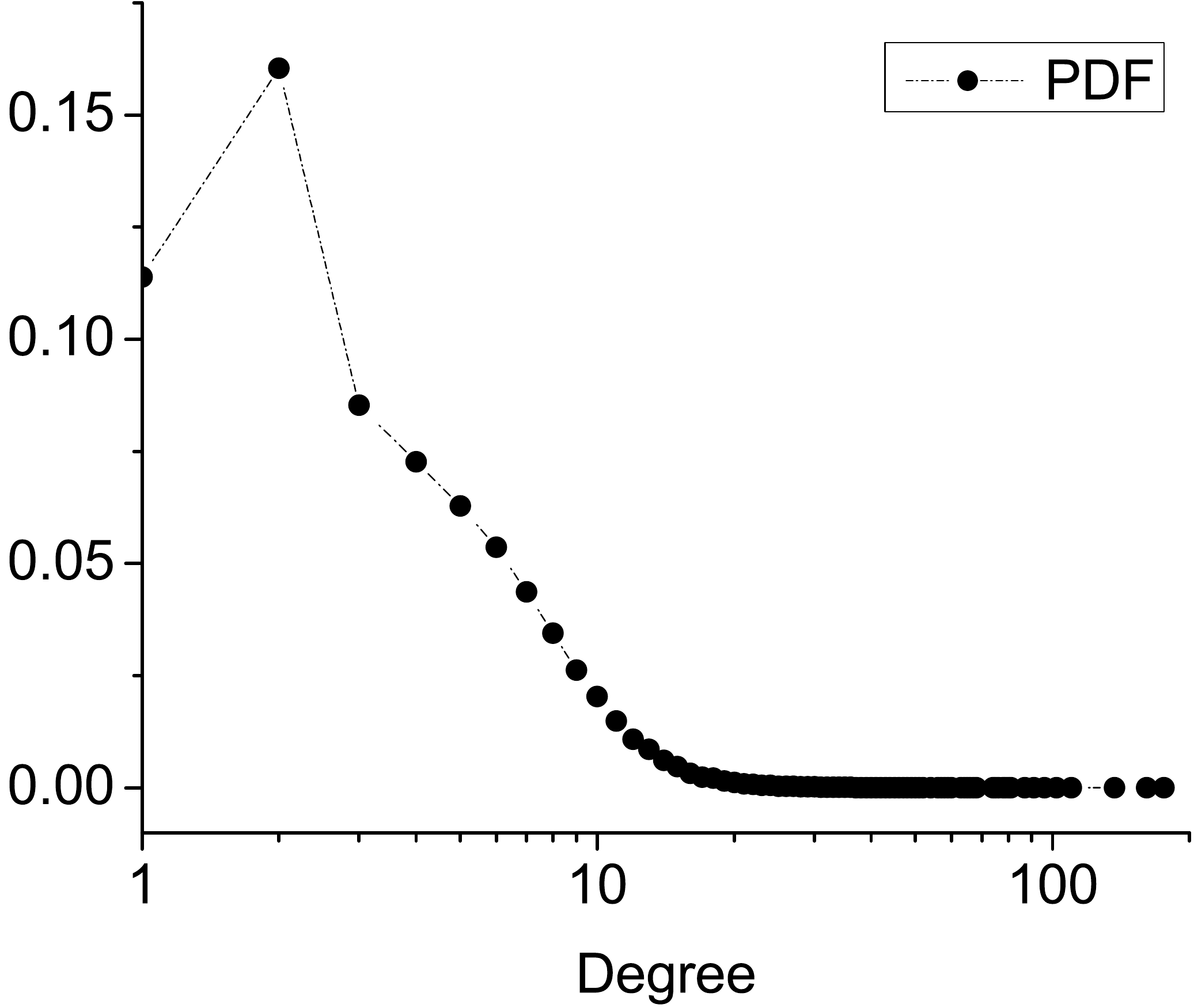}
  }
  \subfigure[Facebook]{
    \includegraphics[width=1.8in]{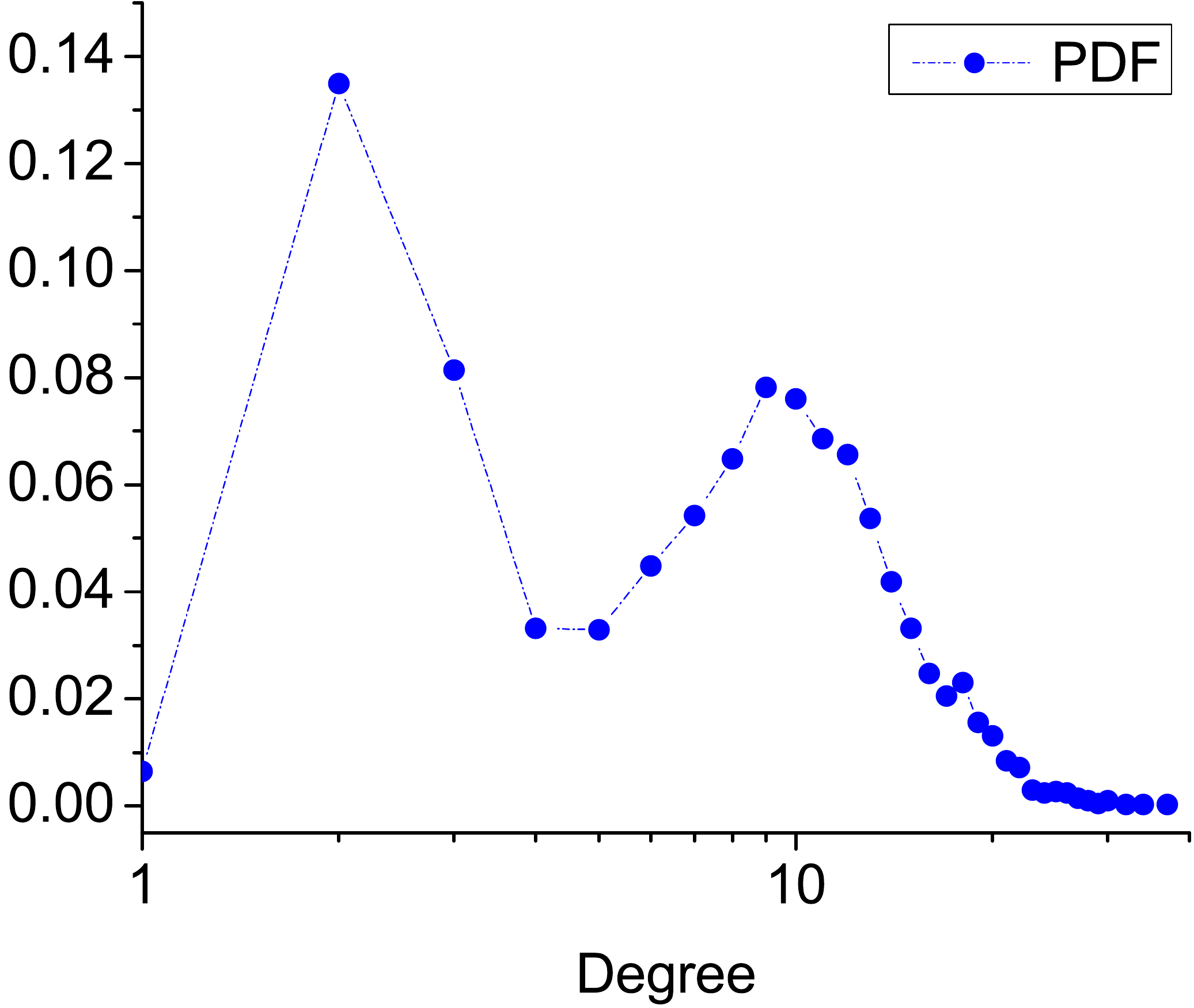}
  }
  \subfigure[Twitter]{
    \includegraphics[width=1.8in]{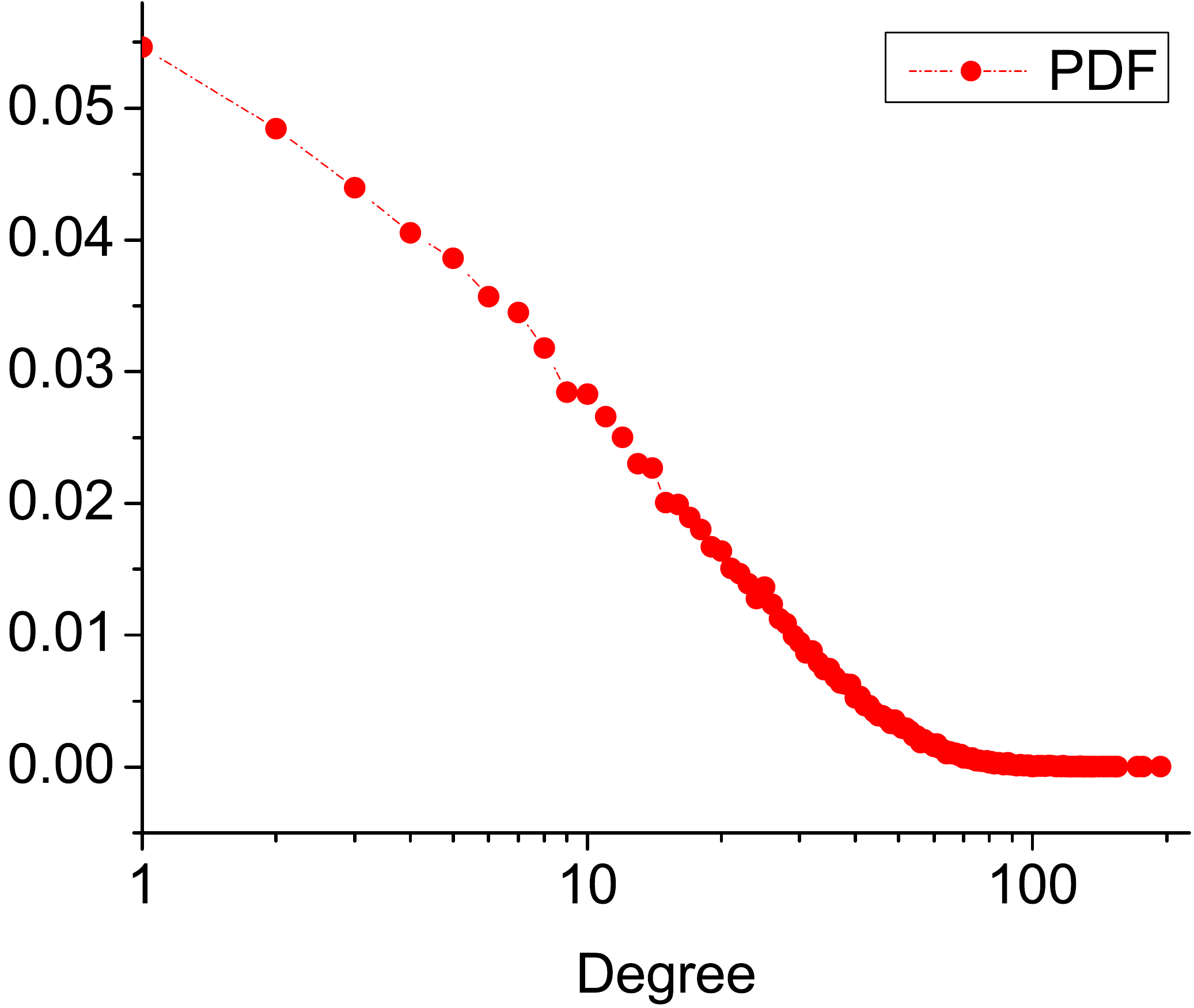}
  }
  \caption{User degree distribution of Google+, Facebook, and Twitter.} \label{f_socialuserdis}
%\vspace{-7mm}
\end{figure*}

We show the user degree distribution of Google+, Facebook, and Twitter
in Fig.\ref{f_socialuserdis}. Basically, the user degree of these three datasets
also shows a power-law-like distribution
(similar to the datasets in Table \ref{tab_data}, the feature degree
of these three datasets show a power-law-like distribution either \cite{report}).
This suggests that the users in the three datasets could be inferred (i.e., de-anonymized here)
based on the associated features.

\subsubsection{FDI Evaluation and Analysis}

\begin{figure}[!tp]
 \centering
  %\subfigure[]{
    \includegraphics[width=3.3in]{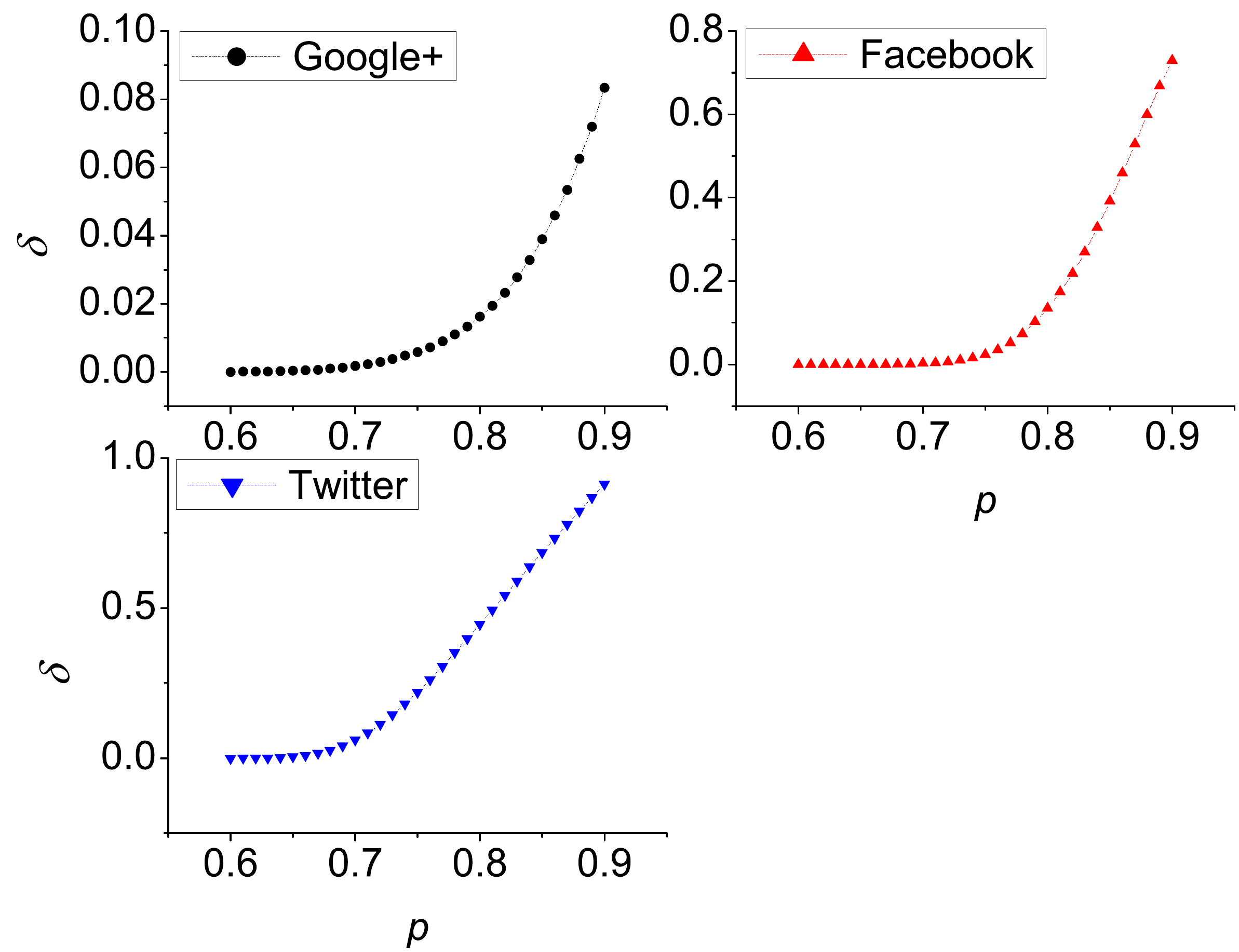}
  %}%\hspace{-8mm}
%  \subfigure[]{
%    \includegraphics[width=2.7in]{figures/traffic-eps-converted-to.pdf}
%  }%\hspace{-8mm}
 % \vspace{-3mm}
  \caption{$(\delta, K)$-inferable verses $p$.
  $K = 10$ in the experiment.} \label{f_dd}
  %\vspace{-6mm}
%\vspace{-8mm}
\end{figure}

To evaluate the FDI of Google+, Facebook and Twitter, we take the same
methodology as in the previous subsection.
We show the FDI of the three datasets in Fig.\ref{f_dd},
where $K = 10$, i.e., we also target the Top-10 inferability of users.
From the result, we have the following observations.

\begin{itemize}
\item
Again, with the increase of $p$, more users become Top-10
inferable in the three datasets. The reason is the same as that in
analyzing Fig.\ref{f_ntd}.

\item
Google+ is much less inferable than that of Facebook and Twitter.
For instance, when $p = 0.8$, $13.47\%$ Facebook users
and $44.65\%$ Twitter users are Top-10 inferable,
while only $1.62\%$ Google+ users are Top-10 inferable.
Even if $p = 0.9$, only $8.34\%$ Google+ users are Top-10 inferable.
This can be explained based on the results in Table \ref{tab_social}
and Fig.\ref{f_socialuserdis}.
First, the user-feature relationship of Google+ is much sparse than
the other two datasets. Second, the degree of most of the Google+
users is very low. Therefore, there is not too much information
can be leveraged to infer the Google+ users.
\end{itemize}

In reality, it is possible to improve the data de-anonymization
performance using more auxiliary information (more features).
Here, our FDI quantification results can provide a benchmark
for evaluating the performance of a data de-anonymization attack.

\section{Discussion} \label{discussion}

In this section, we make more discussion on the proposed FDI quantification
technique, followed by pointing out the future research directions.

\subsection{Theory versus Practice}

Motivated by many existing security and privacy applications,
in this paper, we study the FDI quantification problem.
To the best of our knowledge, we provide the first
FDI quantification technique for general feature-based inference models
from both the distance perspective and the feature distribution perspective.
Using our quantification technique, we also evaluate
the FDI of feature-based network forensics and data de-anonymization applications.

Our quantification is important in several perspectives.
First, our quantification provides the theoretical foundation of many
existing feature-based security and privacy applications,
e.g., network traffic attribution in network forensics \cite{wanwansdm14}\cite{neaperccs14},
linkage attacks and private web search \cite{gershoccs14}\cite{baltrosp12},
and feature-based data de-anonymization \cite{calharusenix15}\cite{afrcalsp14}\cite{narshmsp08}.
Therefore, for such kind of applications, our quantification
closes the gap between the practice and theory.

Second, our quantification can be employed to evaluate the performance
of the existing techniques in the aforementioned security and privacy applications.
Note that, we are aiming to quantify the users who can be inferred
with statistical guarantee based on
their features as well as other users' features.
Meanwhile, we also provide insights on how to design the
inference model (as shown in Section \ref{quantification}).
Therefore, the quantification results
(e.g., the evaluation results in Section \ref{evaluation}) can serve as a benchmark
to evaluate the performance of existing techniques.
For instance, to evaluate the performance of the network traffic attribution system
Kaleido \cite{wanwansdm14}, we can employ the evaluation results in
Section \ref{evaluation} directly\footnote{We can first derive $p$ based on the
training and testing data used in Kaleido. Then, we apply our FDI quantification
to derive the inherent user inferability.
Finally, we can use the theoretical user inferability
to evaluate the performance of Kaleido.
If Kaleido's performance meets the theoretical results, we can conclude
that Kaleido performs well. Otherwise, we can also tell the room for improving
Kaleido.}. Similarly, we can also employ the FDI quantification
to evaluate existing feature-based query linkage attacks, private searching techniques,
data de-anonymization attacks, etc.

Finally, since our quantification can provide a benchmark of existing
feature-based security and privacy applications,
it is evident that
our quantification is helpful for researchers to study and develop
new techniques for these applications.

\subsection{Future Work}

In this paper, we take the first step in understanding the theoretical
foundation of many existing security and privacy applications to the best
of our knowledge. Specifically, we propose the FDI quantification techniques
for distance-based and distribution-based inference models.
There are still several interesting directions to continue the research.
First, it is interesting to further generalize our quantification
to the inference models that take account of both feature distance
and feature distribution.
Second, in addition to the distance/distribution-based models,
it is also meaningful to quantify data's inferability under other
models for more security and privacy applications.
Third, it is an interesting and meaningful direction to develop
some FDI-based evaluation tool which can friendly and conveniently serve the
data inferability analysis for existing feature-based inference-oriented
security and privacy applications.

\section{Related Work} \label{related}

In this section, we survey the related work.
Since we did not have other literature studying the theoretical foundation
or inferability quantification problem for existing
feature-based security and privacy applications to the best of
our knowledge, we focus on briefly summarizing the applications
that our FDI quantification can be applied to.

\emph{User-System Interaction Trace Attribution.}
In \cite{wanwansdm14}, Wang et al. designed a network traffic attribution
system Kaleido. Kaleido leverages a class of inductive discriminant models
to extract user- and context-aware features of network traffic
and then build an efficient inference model to
conduct real time traffic attribution over high-volume network traces.
Another feature-based network forensics application is \cite{neaperccs14},
where Neasbitt et al. proposed ClickMiner, a novel system that aims to
automatically reconstruct user-browser interactions from network traces.
A comprehensive survey on network trace-based forensic frameworks
can be found in \cite{piljosdi10}.

In addition to network traffic-based forensic applications,
there are also many other trace attribution-based security and privacy applications.
For instance, in \cite{berguntissec02}, Bergadano et al. proposed
to employ keystroke dynamics (traces) to perform user authentication;
in \cite{monreisp11}, Monrose et al. designed a technique
to reliably generate a cryptographic key from a user's voice while
speaking a password; and in \cite{zhepalccs11},
Zheng et al. implemented an efficient user verification system
based on mouse movement traces.

\emph{Linkage Attacks and Privacy-preserving Web Search.}
In \cite{gershoccs14}, Gervais proposed a quantitative framework
to understand the web-search privacy given adversary's background knowledge
and attacks.
In \cite{pedsaxpet10}, Peddinti and Saxena analyzed whether query obfuscation
can preserve users' privacy when against an adversarial search engine.
In \cite{jonkumcikm07}, Jones presented attacks to users' query logs
and broke users' privacy.
Recently, Balsa et al. presented a SoK paper on linkage attacks and privacy-preserving
web search \cite{baltrosp12}.

\emph{Feature-based Data De-anonymization.}
In \cite{calharusenix15}, Caliskan-Islam et al. presented a novel
data de-anonymization attack to programmers leveraging the code stylometry.
Afroz et al. presented another stylometry-based de-anonymization attack
in \cite{afrcalsp14}, by which they can identify anonymous authors
of anonymous texts.
In \cite{narshmsp08}, Narayanan and Shmatikov presented a new class of
statistical de-anonymization attacks to high-dimensional micro-data,
e.g., recommendation data, transaction data, and so on.
An off-line de-anonymization attack of bubble forms is presented
in \cite{calclausenix11} by Calandrino et al.

\textbf{Remark.}
In addition to the aforementioned security and privacy applications,
there are also other applications, e.g., feature-based malware detection systems and
intrusion detection systems, that our quantification can be
applicable for analysis.
Although we have many feature-based inference techniques for various
security and privacy applications, their theory foundation is remain unclear.
Furthermore, there is also no theoretical benchmark to evaluate the
performance of existing techniques relative to the inherent performance bound.
To remedy the gap, we conduct the first FDI quantification
in general scenarios from both distance and distribution perspectives.

\section{Conclusion} \label{conclusion}

Considering that many security and privacy applications can be characterized
by the feature-based inference problem, we study the FDI issue in this paper.
First, we conduct the FDI quantification under a naive data model,
under which we demonstrate the conditions to have a desired fraction of
target users to be Top-$K$ inferable.
Subsequently, we extend our quantification to a general data model
by conducting the FDI quantification from both a distance perspective
and a distribution perspective. Our quantification addressed several important yet open problems
and lies the foundation of existing feature-based inference systems/techniques.
Third, based on our quantification, we evaluate the user inferability
in both the network traffic attribution case and the feature-based
data de-anonymization case.
Finally, we point out the implications of this research to
existing feature-based inference systems/tehcniques for
various security and privacy applications.

%\section*{Acknowledgment}
%
%This work was partly supported by NSF-CAREER-CNS-
%******* ******* and DARPA-*********.
%
%%This work was partly supported by NSF-CAREER-CNS-0545667 844144 and DARPA-N10AP20022.

\appendix
\section*{A General Version of Chernoff Bound}
The following version of Chernoff bound applies to
bounded variables with any distribution \cite{chernoff}.
\begin{lemma} \label{l_chernoff}
Let $X_1, X_2, \cdots, X_n$ be random variables such that
$a \leq X_i \leq b$ for all $i$.
Let $X = \sum\limits_{i = 1}^n X_i$
and set $\mu = \mathrm{E}(X)$ (i.e., the expectation value
of $X$). Then, for all $\xi > 0$:
$
\Pr(X \geq (1 + \xi) \mu) \leq \exp(-\frac{2 \xi^2 \mu^2}{n (b - a)^2}),
$
%\begin{align*}
%\Pr(X \geq (1 + \xi) \mu) \leq \exp(-\frac{2 \xi^2 \mu^2}{n (b - a)^2}),
%\end{align*}
and
$
\Pr(X \leq (1 - \xi) \mu) \leq \exp(-\frac{\xi^2 \mu^2}{n (b - a)^2}).
$
%\begin{align*}
%\Pr(X \leq (1 - \xi) \mu) \leq \exp(-\frac{\xi^2 \mu^2}{n (b - a)^2}).
%\end{align*}
\end{lemma}

\end{document}